\documentclass[preprintnumbers,amsmath,amssymb,prd, notitlepage,nofootinbib,twocolumn, superscriptaddress]{revtex4}

\usepackage{graphicx}
\usepackage{dcolumn}
\usepackage{bm}
\usepackage{subfigure}
\usepackage{wasysym}
\usepackage{verbatim}
\usepackage{appendix}
\usepackage{hyperref}
\usepackage{color}
\usepackage[normalem]{ulem}
\usepackage{capt-of}

\newcommand{\Mnu}{$\Sigma m_{\nu}$}
\newcommand{\Nnu}{$N_{\rm eff}$}

\hypersetup{colorlinks=true}

\newcommand{\LCDM}{$\mathrm{\Lambda}$CDM}
\newcommand{\geff}{$\textit{G}_{\rm eff}$}
\newcommand{\loggeff}{$\rm{log_{10}}(\textit{G}_{\rm{eff}}{\rm MeV}^2)$}

\begin{document}

\title{The Atacama Cosmology Telescope: The Persistence of Neutrino Self-Interaction in Cosmological Measurements}

\author{Christina D. Kreisch} 
\affiliation{Department of Astrophysical Sciences, Princeton University, Princeton, New Jersey 08544, USA}
\author{Minsu Park}
\affiliation{Center for Particle Cosmology, Department of Physics and Astronomy,
University of Pennsylvania, Philadelphia, Pennsylvania 19104, USA}
\author{Erminia Calabrese}
\affiliation{School of Physics and Astronomy, Cardiff University, The Parade, Cardiff, CF24 3AA, UK}
\author{Francis-Yan Cyr-Racine}
\affiliation{Department of Physics and Astronomy, University of New Mexico, 210 Yale Blvd NE, Albuquerque, New Mexico 87106, USA}

\author{Rui An}
\affiliation{Department of Physics \& Astronomy, University of Southern California, Los Angeles, CA, 90007, USA}

\author{J. Richard Bond}
\affiliation{Canadian Institute for Theoretical Astrophysics, University of Toronto, 60 St. George St., Toronto, ON M5S 3H8, Canada}

\author{Olivier Dor\'e}
\affiliation{Jet Propulsion Laboratory, California Institute of Technology, Pasadena, California 91109, USA}
\affiliation{California Institute of Technology, Pasadena, California 91125, USA}
\author{Jo Dunkley}
\affiliation{Joseph Henry Laboratories of Physics, Jadwin Hall, Princeton University, Princeton, NJ 08544}
\affiliation{Department of Astrophysical Sciences, Princeton University, Princeton, New Jersey 08544, USA}

\author{Patricio Gallardo}
\affiliation{Kavli Institute for Cosmological Physics, University of Chicago, Chicago, IL 60637, USA}

\author{Vera Gluscevic}
\affiliation{Department of Physics \& Astronomy, University of Southern California, Los Angeles, CA, 90007, USA}

\author{J.~Colin Hill}
\affiliation{Department of Physics, Columbia University, New York, NY, USA 10027}
\affiliation{Center for Computational Astrophysics, Flatiron Institute, New York, NY 10010, USA}

\author{Adam D.~Hincks}
\affiliation{David A. Dunlap Department of Astronomy \& Astrophysics, University of Toronto, 50 St. George St., Toronto, ON M5S 3H4, Canada}
\affiliation{Specola Vaticana (Vatican Observatory), V-00120 Vatican City State}

\author{Mathew S.~Madhavacheril}
\affiliation{Perimeter Institute for Theoretical Physics, 31 Caroline Street N, Waterloo ON N2L 2Y5, Canada}

\author{Jeff McMahon}
\affiliation{Department of Astronomy and Astrophysics, University of Chicago, Chicago, IL 60637, USA}
\affiliation{Kavli Institute for Cosmological Physics, University of Chicago, Chicago, IL 60637, USA}
\affiliation{Department of Physics, University of Chicago, Chicago, IL 60637, USA}
\affiliation{Enrico Fermi Institute, University of Chicago, Chicago, IL 60637, USA}

\author{Kavilan Moodley}
\affiliation{Astrophysics Research Centre, University of KwaZulu-Natal, Westville Campus, Durban 4041, South Africa}
\affiliation{School of Mathematics, Statistics \& Computer Science, University of KwaZulu-Natal, Westville Campus, Durban 4041, South Africa}

\author{Thomas~W.~Morris}
\affiliation{Joseph Henry Laboratories of Physics, Jadwin Hall, Princeton University, Princeton, NJ 08544}

\author{Federico Nati}
\affiliation{Department of Physics, University of Milano-Bicocca, Piazza della Scienza 3, 20126 Milano (MI), Italy}

\author{Lyman A. Page}
\affiliation{Joseph Henry Laboratories of Physics, Jadwin Hall, Princeton University, Princeton, NJ 08544}

\author{Bruce Partridge}
\affiliation{Department of Astronomy, Haverford College, Haverford, PA 19041, USA}

\author{Maria Salatino}
\affiliation{Stanford University, Stanford, CA 94305, USA}
\affiliation{Kavli Institute for Particle Astrophysics and Cosmology, Stanford, CA 94305, USA}

\author{Crist\'obal Sif\'on}
\affiliation{Instituto de F\'isica, Pontificia Universidad Cat\'olica de Valpara\'iso, Casilla 4059, Valpara\'iso, Chile}

\author{David N.~Spergel}
\affiliation{Center for Computational Astrophysics, Flatiron Institute, New York, NY 10010, USA}
\affiliation{Department of Astrophysical Sciences, Princeton University, Princeton, New Jersey 08544, USA}

\author{Cristian Vargas}
\affiliation{Instituto de Astrof\'isica and Centro de Astro-Ingenier\'ia, Facultad de F\'isica, Pontificia Universidad Cat\'olica de Chile, Av. Vicu\~na Mackenna 4860, 7820436 Macul, Santiago, Chile}

\author{Edward J.~Wollack}
\affiliation{NASA Goddard Space Flight Center, 8800 Greenbelt Rd, Greenbelt, MD 20771, USA}

\date{\today}

\begin{abstract}
 
We use data from the Atacama Cosmology Telescope (ACT) DR4 to search for the presence of neutrino self-interaction in the cosmic microwave background. Consistent with prior works, the posterior distributions we find are bimodal, with one mode consistent with \LCDM\ and one where neutrinos strongly self-interact. By combining ACT data with large-scale information from \textit{WMAP}, we find that a delayed onset of neutrino free streaming caused by significantly strong neutrino self-interaction is compatible with these data at the $2-3\sigma$ level.
As seen in the past, the preference shifts to \LCDM\ with the inclusion of \textit{Planck} data. We determine that the preference for strong neutrino self-interaction is largely driven by angular scales corresponding to $700 \lesssim \ell \lesssim 1000$ in the ACT E-mode polarization data. This region is expected to be key to discriminate between neutrino self-interacting modes and will soon be probed with more sensitive data. 

\end{abstract}
\maketitle

\section{Introduction}

Neutrinos remain an elusive component of the Standard Models of particle physics and cosmology. While cosmological measurements have placed some of the strongest constraints on the sum of neutrino masses  (see e.g.,~Refs.~\cite{Planck:2018vyg,Lattanzi:2017ubx}), we do not yet know the value. The precise mechanism for the generation of such neutrino masses is also still uncertain. Further, the presence of anomalies in terrestrial neutrino experiments~\cite{Aguilar-Arevalo:2018gpe,Aguilar:2001ty,Aartsen:2017bap,TheIceCube:2016oqi,Adamson:2017zcg,Adamson:2017uda,Dentler:2018sju} may indicate, if confirmed, that yet unknown physics exists in the neutrino sector, hence providing a window into physics beyond the Standard Model.

In particular, new physics altering the free-streaming nature of neutrinos in the early universe has received renewed interest in recent years 
(see e.g.,~Refs.~\cite{Choi:2018gho,song2018,Lorenz:2018fzb,Barenboim:2019tux,Forastieri:2019cuf,Smirnov:2019cae,Escudero:2019gvw,Ghosh:2019tab,Funcke:2019grs,Sakstein:2019fmf,Mazumdar:2019tbm,Blinov:2020hmc,deGouvea:2019qaz,Froustey:2020mcq,Babu:2019iml,Deppisch:2020sqh,Kelly:2020pcy,EscuderoAbenza:2020cmq,He:2020zns,Ding:2020yen,Berbig:2020wve,Gogoi:2020qif,Barenboim:2020dmg,Das:2020xke,Mazumdar:2020ibx,Brinckmann:2020bcn,Kelly:2020aks,Esteban:2021ozz,Arias-Aragon:2020qip,Du:2021idh,CarrilloGonzalez:2020oac,Huang:2021dba,Sung:2021swd,Escudero:2021rfi,RoyChoudhury:2020dmd,Carpio:2021jhu,Orlofsky:2021mmy,Esteban21,Venzor22}). 
In the Standard Model, neutrinos decouple from the primordial plasma and begin to free stream when the universe has cooled to a temperature of $\sim$1.5$~\mathrm{MeV}$. While freely streaming, neutrinos still interact gravitationally with the rest of the universe, tugging on any particles in their paths while they pass by. Observationally, this damps the amplitude of photon fluctuations and shifts them to slightly larger scales~\cite{Bashinsky:2003tk,Hou:2011ec,Follin:2015hya}, impacting the amplitude and phase of the observed cosmic microwave background (CMB) temperature and polarization power spectra.

Introducing new physics in the neutrino sector by allowing them to self-interact can, however, significantly delay the time at which neutrinos begin to free-stream. Such a delay abates how long neutrinos gravitationally tug on the photons, leaving a measurable imprint on the CMB~\cite{Cyr-Racine:2013aa,Baumann15}. The delay in free-streaming also impacts the evolution of the two Newtonian gravitational potentials $\phi$ and $\psi$, leading to scale-dependent effects on the growth of matter fluctuations. See Ref.~\cite{kreisch} for a thorough discussion of these effects. Through the combination of effects, neutrino self-interactions can be constrained with CMB measurements, baryon acoustic oscillation (BAO) measurements, and other large scale structure (LSS) measurements (see e.g.,~Refs.~\cite{Cyr-Racine:2013aa,Archidiacono:2013dua,lancaster,oldengott17,kreisch,Barenboim:2019tux, park19,Das:2020xke,Mazumdar:2020ibx,RoyChoudhury:2020dmd}).  

In its simplest implementation,  self-interactions can be described by an effective four-fermion interaction parameterized by a dimensionful Fermi-like constant $G_{\rm eff}$~\cite{Ng:2014pca}. This effective coupling constant determines the neutrino self-interaction rate, $\Gamma_\nu \propto G_{\rm eff}^2 {T}_\nu^5$ where ${T}_\nu$ is the homogeneous temperature of the neutrino bath. Since this interaction only happens between neutrinos, it does not alter the physics and timing of neutrinos decoupling from the rest of the primordial plasma, as discussed in Ref.~\cite{Grohs:2020xxd}. We note however that any ultraviolet completion of the effective four-fermion interaction is subject to several strong constraints, including from supernovae~\cite{Kolb:1987qy,Manohar:1987ec,Dicus:1988jh,Davoudiasl:2005fd,Sher:2011mx,Fayet:2006sa,Choi:1989hi,Blennow:2008er,Galais:2011jh,Kachelriess:2000qc,Farzan:2002wx,Zhou:2011rc,Jeong:2018yts,Chang22}, big bang nucleosynthesis~\cite{Ahlgren:2013wba,Huang:2017egl,Venzor:2020ova}, neutrino observations with the IceCube experiment ~\cite{Ng:2014pca,Ioka:2014kca,Cherry:2016jol}, particle colliders ~\cite{Bilenky:1992xn,Bardin:1970wq,Bilenky:1999dn,Brdar:2020nbj,Lyu:2020lps}, and those arising from meson, leptons, tritium, and gauge boson decay kinematics~\cite{Lessa:2007up,Bakhti:2017jhm,Arcadi:2018xdd,Brdar:2020nbj,Blinov:2019gcj,Lyu:2020lps}. Taken literally, these bounds exclude values of $G_{\rm eff}$ large enough to affect cosmological observables. Therefore, the  $G_{\rm eff}$ parameterization used in this work should not be interpreted as an actual particle model of neutrino self-interaction, but rather as a proxy controlling the onset of neutrino free-streaming in our universe. 

With this in mind, previous works~\cite{Cyr-Racine:2013aa, lancaster, kreisch, Archidiacono:2013dua, oldengott17, Barenboim:2019tux, park19,Das:2020xke,Mazumdar:2020ibx,Brinckmann:2020bcn,RoyChoudhury:2020dmd} have interestingly found that an effective neutrino self-interaction strength orders of magnitude larger than the standard electroweak interaction can be compatible with CMB and BAO data. Unlike other popular cosmological extensions, the $G_{\rm eff}$ posterior probability distribution is characterized by two distinct islands in parameter space: a strongly interacting mode, SI$\nu$, with $G_{\rm eff}\sim 10^{-1.5}\,\mathrm{MeV}^{-2}$, and a moderately interacting mode, MI$\nu$, with $G_{\rm eff}\sim10^{-4}\,\mathrm{MeV}^{-2}$  that is nearly indistinguishable from $\Lambda \mathrm{CDM}$. This bimodality stems from a multi-parameter degeneracy with $G_{\rm eff}$~\cite{lancaster} involving the angular size of the baryon-photon sound horizon at last scattering $\theta_*$, the amplitude of scalar fluctuations $A_{\rm s}$, and the scalar spectral index $n_{\rm s}$. The strong neutrino self-interactions of the SI$\nu$ mode shift the phase and boost the amplitude of the multipoles entering the causal horizon before the onset of neutrino free-streaming. To reconcile these effects with cosmological data, larger $\theta_*$ and lower $n_{\rm s}$ values are preferred when $G_{\rm eff}$ is large, which in turns result in a lower value of $A_{\rm s}$ to ensure consistency with low CMB multipoles. On the other hand, the MI$\nu$ mode is characterized by  values of $\theta_*$, $A_{\rm s}$, and $n_{\rm s}$ approximately equal to their $\Lambda \mathrm{CDM}$ values.

The recent special interest for this model arises from the fact that the SI$\nu$ mode belongs to the family of scenarios that bring the CMB and Cepheid-calibrated SNIa measurements of the Hubble constant, $H_0$, closer by introducing a new species relevant in the early universe to reduce the sound horizon at recombination (see e.g.,~Refs.~\cite{Verde:2019ivm,DiValentino:2021izs}). Preferring a higher value of $N_{\rm eff}$, the SI$\nu$ mode is coincident with larger values of $H_0$ and lower values of $\sigma_8$, the amplitude of linear density fluctuations at 8 $h^{-1}\mathrm{Mpc}$, offering a potential simultaneous resolution to both the $\sigma_8$ tension and discrepancies in $H_0$. 

Nevertheless, as discussed in Refs.~\cite{kreisch,Das:2020xke,Mazumdar:2020ibx,Brinckmann:2020bcn,RoyChoudhury:2020dmd}, the inclusion of the \textit{Planck} CMB polarization data~\cite{Planck:2018vyg} disfavors the SI$\nu$ mode compared to the MI$\nu$, casting serious doubt on the viability of this mechanism to resolve the current tensions (see Ref.~\cite{Cyr-Racine:2021oal} for a more detailed explanation of this limitation). 
Despite being statistically suppressed, the SI$\nu$ mode is not entirely ruled out by these analyses. This then asks the question of what kind of cosmological data \emph{could} eliminate the viability of a late onset of neutrino free streaming.

The low value of the spectral index $n_{\rm s}$ associated with the SI$\nu$ mode provides an important clue: cosmological data probing a broad range of scales can provide an important lever arm to constrain the spectral tilt and detect any deviation from its $\Lambda$CDM value. High-resolution observations of the CMB temperature and polarization spectra probing small angular scales inaccessible to the \textit{Planck} satellite are a promising candidate for such an observational constraint. In this work, we use high-resolution CMB data from four observing seasons of the Atacama Cosmology Telescope (ACT)~\cite{actlike1} to probe neutrino self-interaction in the early universe, which is a step towards higher sensitivity measurements from the complete ACT dataset, the Simons Observatory~\cite{SimonsObservatory:2018koc} and CMB-S4~\cite{cmbs4}.

Previous works have shown the compatibility of  ACT measurements with other models increasing the sound horizon at recombination, such as early dark energy (EDE) and pseudoscalar sterile neutrino self-interactions~\cite{colinede,sterileint}. Though the physics of how they increase the inferred $H_0$ from the CMB is similar, the underlying physics and the subsequent perturbation theory and phenomenology can be vastly different, thus motivating further exploration of these class of models with a wide variety of future data-sets.

We show below that a delayed onset of neutrino free streaming brought on by significant neutrino self interactions still appears compatible with CMB observations at small angular scales from ACT. This is the case for both ACT alone and in combination with data from the Wilkinson Microwave Anisotropy Probe (\textit{WMAP})~\cite{Hinshaw:2012aka,bennett/etal:2013}.


The paper is organized as follows. In \autoref{sec:data}, we present the cosmological models, data, parameter choices, and statistical tools used in our analyses. Our main results are presented in \autoref{sec:survive}. In \autoref{sec:Emode} we highlight the importance of the ACT E-mode polarization for our results. We consider the impact of BAO measurements on our results in \autoref{sec:BAO_H0} and briefly discuss our systematic tests in \autoref{sec:syst}. We conclude in \autoref{sec:conc}.

\section{Data \& Methodology}
\label{sec:data}

\subsection{Models, Data, and Parameter Choices}
Our baseline cosmological model includes three massive neutrinos with degenerate masses that can scatter among themselves with an interaction rate $\Gamma_\nu \propto G_{\rm eff}^2 {T}_\nu^5$. This parallels the analysis done by \textit{Planck}~\cite{Planck:2018vyg}.\footnote{Our analysis here differs from that presented in Ref.~\cite{kreisch}, which used a single massive neutrinos containing all the mass, in addition to massless neutrinos.} The details of the Boltzmann equations involving such massive self-interacting neutrinos are provided in Ref.~\cite{kreisch} (see also Ref.~\cite{oldengott15}). Within this baseline model, the parameters $N_{\rm eff}$ (which is used to adjust the neutrino temperature $T_\nu$) and the sum of neutrino masses $\sum m_\nu$ are also allowed to vary freely from their standard values of 3.046 and 0.06eV respectively. Our baseline model is thus described by a total of 9 parameters once the six standard $\Lambda$CDM parameters\footnote{Baryon density $\Omega_b h^2$, cold dark matter density $\Omega_c h^2$, angular peak position $\theta_s$, spectral index $n_s$ and amplitude $A_s$, and optical depth to reionization $\tau$.} are included. This model is denoted as ``$G_{\rm eff} + N_{\rm eff} + \sum m_\nu$'' in what follows. We also consider a simpler extension of $\Lambda$CDM in which only the neutrino interaction strength $G_{\rm eff}$ is added, with $N_{\rm eff}$ and $\sum m_\nu$ fixed at their standard values. This model is simply referred to as ``${\rm G}_{\rm eff}$''. 

Throughout our analysis, we used modified versions of the codes CAMB\footnote{\url{https://github.com/ckreisch/IntNuCAMB}}~\cite{CAMB} and CosmoMC+Multinest~\cite{cosmomc,multinest}, equivalent to those used in Ref.~\cite{kreisch}. This analysis framework assumes a linear evolution of perturbations. Although negligible at the scales probed by \textit{Planck}, at smaller scales like those measured by ACT, non-linear gravitational lensing effects impact \Mnu\ and \Nnu\ estimates and therefore it is likely that they will also alter estimates of \geff. However, we will show later that, despite entering the non-linear regime for lensing,  multipoles above $2500$ contribute negligible information to the constraint and therefore a linear analysis here is valid. For future analyses with even stronger influence from $\ell > 2500$, analyzing neutrino self-interactions in the non-linear regime may become necessary.

We use uniform priors on all parameters, except for the \textit{Planck} {absolute map-level calibration parameter by the square which all \textit{Planck} spectra are divided} which has the Gaussian prior $y_{\rm cal} = 1.0000 \pm 0.0025$, and the optical depth to reionization which has the Gaussian prior $\tau = 0.065 \pm 0.015$ (see Ref.~\cite{actlike1}) and replaces low-$\ell$ polarization data. As in previous work, we place a uniform prior on the logarithm of the coupling constant $G_{\rm eff}$, but we here extend the lower range of the prior to $-8.0$ to include smaller coupling values and allow for any shift in the MI$\nu$ mode location. 
We also utilize nested sampling~\cite{skilling2006} to thoroughly sample the multi-modal posteriors. For this we use 2000 live points, set the target sampling efficiency to 0.3, set the accuracy threshold on the log Bayesian evidence to 20\% to ensure the accuracy of credible intervals. We refer to `modes' as disjoint regions of parameter space that isolate the islands of the multi-modal posterior distribution. We utilize the mode separation feature in the Multinest algorithm~\cite{multinest} to isolate each mode and compute the parameter posterior distributions for that mode. 

In this work we present results on neutrino self-interactions in the presence of the ACT DR4 data. We further combine ACT with \textit{WMAP} and \textit{Planck} CMB data, as well as a selection of low-redshift datasets. We denote the data as follows:

\begin{itemize}
    \item ACT: ACT CMB {\tt actpollite\_dr4} likelihood for observing seasons 2013-2016 (DR4), containing TT (temperature autocorrelation) measurements spanning $600<\ell<4126$ and TE (temperature-polarization correlation) and EE (polarization-autocorrelation) measurements spanning $350<\ell<4126$~\cite{actlike2}. 
    \item \textit{Planck}: \textit{Planck} 2018 CMB {\tt plik\_lite} high-$\ell$ likelihood, containing TT measurements spanning $30<\ell<2508$ and TE/EE measurements spanning $30<\ell<1996$, as well as the {\tt commander} low-$\ell$ TT likelihood, with measurements spanning $2<\ell < 29$~\cite{plancklike2018}. 
    \item \textit{WMAP}: \textit{WMAP} CMB 9-year observations, containing TT measurements spanning $2 < \ell < 1200$ and TE measurements spanning $24 < \ell < 800$~\cite{Hinshaw:2012aka,bennett/etal:2013}.
    \item BAO: Baryon acoustic oscillation (BAO) measurements from Sloan Digital Sky Survey Main Galaxy Sample, Six-degree Field Galaxy Survey, and Data Release 12~\cite{BAO1,BAO2,BAO3}. 
    \item Lens: \textit{Planck} CMB lensing data from the 2018 release containing lensing multipoles $8\leq L\leq 400$~\cite{plancklens:2018}. This is only included when also adding BAO to combinations with \textit{Planck}. Thus, we do not make note of the lensing measurement when labeling the data combination.
\end{itemize}

When combining ACT with \textit{Planck} or \textit{WMAP}, we follow the procedure described in Ref.~\cite{actlike1}, i.e., we remove ACT TT data below $\ell<1800$ in analyses with \textit{Planck} while we use the full ACT dataset in combination with \textit{WMAP}. 

\subsection{Model-Comparison Tools}

We use a variety of techniques to assess the statistical significance of the models we consider. To compare statistically distinct modes of the posterior, we compute their Bayes factor
\begin{align}
    \mathcal{B}_{\mathrm{SI}\nu} \equiv  \frac{\mathcal{Z}_{\mathrm{SI}\nu}}{\mathcal{Z}_{\mathrm{MI}\nu}}
\end{align}
which compares the modes' Bayesian evidence $\mathcal{Z}$, defined as the parameter-averaged likelihood of the data,
\begin{align}
  \mathcal{Z}_j \equiv \mathrm{Pr}\left(\mathbf{d}|\mathcal{M}_j\right) = \int_{\Omega_{\theta}}\mathrm{Pr}\left(\mathbf{d}|\mathbf{\theta},\mathcal{M}_j\right)\mathrm{Pr}\left(\mathbf{\theta}|\mathcal{M}_j\right)d\mathbf{\theta},
\end{align}
where $\textbf{d}$ is the data, $\mathbf{\theta}$ are the parameters describing model $\mathcal{M}$, $\mathcal{M}_j$ denotes the $j$-th mode of the posterior distribution on the space of $\mathbf{\theta}$, and $\Omega_{\theta}$ is the entire parameter space. {Note that we are using the Bayes factor to compare the modes' statistical significances, instead of model comparison.~\cite{lancaster}}

We also assess the relative statistical significance between two modes of the posterior by computing their maximum likelihood ratio:
\begin{align}
  \mathcal{R}_{\mathrm{SI}\nu} = \frac{\mathrm{max}\left[\mathcal{L}\left(\theta_{\mathrm{SI}\nu}|\mathbf{d}\right)\right]}{\mathrm{max}\left[\mathcal{L}\left(\theta_{\mathrm{MI}\nu}|\mathbf{d}\right)\right]},
\end{align}
where $\theta_{\mathrm{SI}\nu}$ and $\theta_{\mathrm{MI}\nu}$ are the parameters describing the SI$\nu$ and MI$\nu$ modes' best-fit parameters, respectively.

To assess how well the different models fit the data, we also compute $\Delta \chi^2$ values. We then translate these values into a significance in terms of Gaussian standard deviations (i.e.~$\sigma$s). We assume that $\Delta \chi^2$ is distributed according to the $\chi^2-$distribution. When comparing fits across different numbers of free parameters, we assume the $\chi^2-$distribution with $k$ degrees of freedom where $k$ is the difference in parameter numbers. This effectively penalizes the additional parameters. Then we find the $\sigma$ (Gaussian significance) whose C.L. matches the cumulative probability function at the given $\Delta \chi^2$, following the same method as in \cite{colinede}. We note, however, that the multi-dimensional posteriors may not be sufficiently Gaussian for either mode (especially MI$\nu$), so Gaussian significances should be interpreted with caution. 

We finally compute the Akaike information criterion (AIC)~\citep{AIC} to penalize the extra parameters added to the interacting neutrino model. The $\Delta \mathrm{AIC}$ between two models is computed as: 
\begin{align}
\Delta \mathrm{AIC} = \mathrm{AIC}_{\mathrm{I}\nu} - \mathrm{AIC}_{\Lambda\mathrm{CDM}} = \Delta \chi^2 + 2\Delta k.
\end{align}
A lower AIC value provides a better fit, so a negative $\Delta \mathrm{AIC}$ value indicates preference for interacting neutrinos.  

\section{CMB-only Results with emphasis on ACT}\label{sec:survive}

\begin{figure}
\begin{center}
\includegraphics[width=0.5\textwidth]{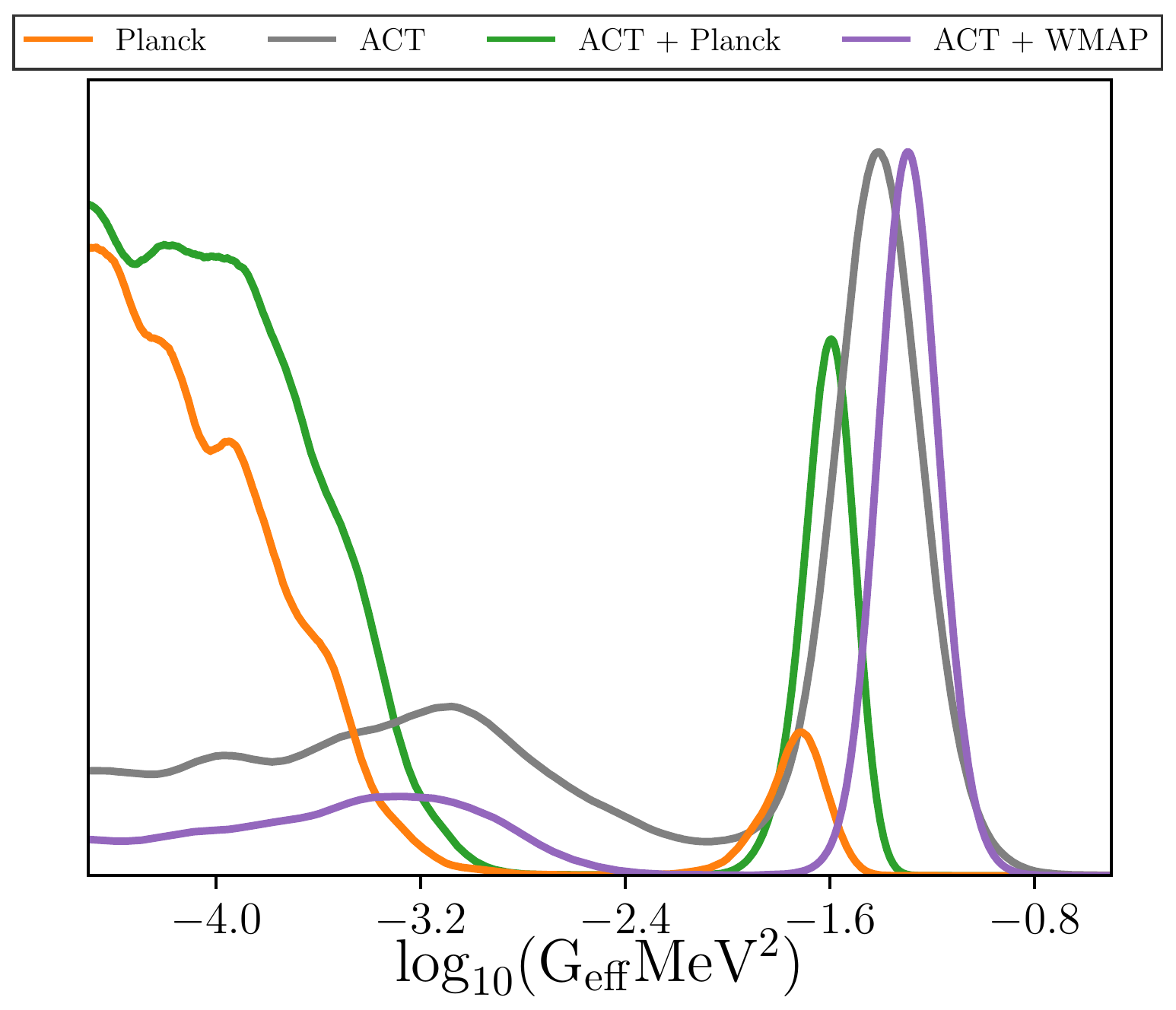}
\caption{Marginalized posterior distribution of  $G_{\rm eff}$ for the different CMB datasets for our baseline $G_{\rm eff} + N_{\rm eff} + \sum m_\nu$ model. We limit the range of the horizontal axis displayed here (despite the prior extending farther in either direction) to better show the details of the two posterior modes. We minimally smooth the posteriors in order to maintain the bimodal features. As such, small values of $G_{\rm eff}$ can appear unconverged, as indicated by the wiggly lines for low $G_{\rm eff}$.}
\label{fig:posteriorsgeff}
\end{center}
\end{figure}

\begin{table}[t!]
  \caption{Mode comparison for our baseline $G_{\rm eff}$+$N_{\rm eff}$+$\sum m_\nu$ model for different CMB dataset combinations. $\mathcal{B}_{\mathrm{SI}\nu}$ is the Bayes factor between the SI$\nu$ and the MI$\nu$ modes, $\mathcal{R}_{\mathrm{SI}\nu}$ is the maximum likelihood ratio, and $\Delta \chi^2_{\rm Tot} = \chi^2_{\mathrm{SI}\nu} - \chi^2_{\mathrm{MI}\nu}$.   \label{tab:ev_likeratios}}
  \begin{ruledtabular}
  \begin{tabular}{ccccccc}
   & ACT & \textit{Planck} & ACT + \textit{WMAP} & ACT + \textit{Planck} \\ 
  \hline \\ [-2ex]
  $\mathcal{B}_{\mathrm{SI}\nu}$ 	& $1.5 \pm  0.2$	& $0.01  \pm 0.01$	& $2.8  \pm  0.6$ & $0.05  \pm 0.02$ \\
  $\mathcal{R}_{\mathrm{SI}\nu}$ 	& $2.6$	& $0.4$	& $6.5$ & $1.1$ \\
  $\Delta\chi^2_{\rm Tot}$ & $-1.9$	& $1.8$	& $-3.7$ & $-0.2$ \\ [0.5ex]
  \end{tabular}
\end{ruledtabular}
\end{table}


The parameter constraints on our baseline $G_{\rm eff}$+$N_{\rm eff}$+$\sum m_\nu$ model using various combinations of ACT, \textit{WMAP}, and \textit{Planck} data are shown in \autoref{fig:posteriorsgeff}. ACT data, both alone and when combined with \textit{WMAP}, show some preference for strong neutrino self-interactions and the resulting delayed onset of neutrino free streaming. The distribution of \geff\ is bimodal for all dataset combinations presented, but the strongly interacting mode is relatively preferred with the ACT data alone as well as the ACT data combined with \textit{WMAP} data, whereas the weakly-interacting mode, compatible with \LCDM, is preferred by the \textit{Planck} data. Adding ACT data to \textit{Planck} increases the probability of the $\mathrm{SI}\nu$ mode compared to \textit{Planck} alone, but keeps an overall preference for the $\mathrm{MI}\nu$ mode. It is interesting to note that the ACT+\textit{WMAP} combination, which together probe a broad range of angular scales, comparatively to \textit{Planck} favors a late onset of neutrino free streaming, although $\Lambda$CDM still provides a good fit to the data. The improvement in overall $\chi^2$ for this interacting model, compared to $\Lambda$CDM, is 13.7 for three extra parameters for ACT+\textit{WMAP}, which we translate to a 2.9$\sigma$ preference. For ACT+\textit{Planck} the improvement in $\chi^2$ is only 1.7 for the $\mathrm{SI}\nu$ mode, with $\Lambda$CDM preferred. The physical reasons driving these preferences and differences will be discussed in the next few sections.

In \autoref{tab:ev_likeratios} we quantify the relative significance of the SI$\nu$ and MI$\nu$ modes within our baseline model. For ACT-only, a Bayes factor of $1.5\pm0.2$ implies a minor preference for strong self interaction and is consistent with noise.\footnote{We note that Multinest was not able to automatically separate the modes. To compare evidence, we ran two separate chains with a prior on \loggeff\ of $[-4.0,-2.0]$ and $[-2.0,0.0]$, respectively, and with equivalent priors for all other parameters. We chose these $G_{\rm eff}$ priors in order to maintain the same prior volume between the two modes, and we safely are able to cut at $-4$ instead of $-8$ since the ACT-only 1D \geff\ posterior drops off, as in \autoref{fig:posteriorsgeff}.} With \textit{WMAP} added, the Bayes factor grows to $2.8\pm0.6$, corresponding to an increased preference for SI$\nu$ ~\cite[see Table 4 in Ref.][where the inverse corresponds to our definition of the Bayes factor]{bayes}. \textit{Planck} alone and ACT+\textit{Planck} show values below 1, preferring an early onset of neutrino free streaming consistent with $\Lambda \mathrm{CDM}$. Similar conclusions can be drawn from the maximum likelihood ratios. 

Credible intervals for the different cosmological parameters within our baseline model are given in Appendix~\ref{app:params} in \autoref{tab:bestfit_sinu_baseline} for the SI$\nu$ mode and in \autoref{tab:bestfit_minu_baseline} for the MI$\nu$ mode. The estimated Hubble constant for these models are lower than for \LCDM, due to marginalizing over a broad range of neutrino masses.\\

{\bf The Efficacy of $\mathrm{G}_{\rm eff}$ alone.} To explore how much of the slight preference for the SI$\nu$ mode we see when ACT is involved is influenced by some movements in other neutrino parameters degenerate with \geff, we test a simplified model in which we fix $N_{\rm eff} = 3.046$ and $\sum m_\nu=0.06$ eV, letting only $G_{\rm eff}$ and the standard six $\Lambda$CDM parameters vary freely. We find that while removing the $N_{\rm eff}$ and $\sum m_\nu$ freedom has little impact on the SI$\nu$ mode for ACT+\textit{WMAP}, it does significantly suppress the existence of the MI$\nu$ mode for this data combination. This indicates that the relevance of the MI$\nu$ mode depends on the values recovered for other neutrino parameters in the fit, while the existence of the SI$\nu$ mode is not driven by the specific values of $N_{\rm eff}$ and $\sum m_\nu$. In other words, the delayed onset of neutrino free-streaming caused by a large value of $G_{\rm eff}$ is key to ACT's preference for the SI$\nu$ mode. 
Indeed, we find that adding the single $G_{\rm eff}$ parameter improves the fit to ACT data moderately more than other simple extensions considered in the literature or more than the combination $N_{\rm eff}$+$\sum m_\nu$ -- see  \autoref{tab:bestfit_sinu_geff} and \autoref{tab:chi2lcdm_awext} in Appendix~\ref{app:params}. Here the $\rm{G_{eff}}$ model slightly increases $H_0$ compared to the \LCDM\ result, with $H_0=69.3\pm1.1$ versus $67.6\pm1.1$~km/s/Mpc for \LCDM\ \cite{actlike1}. \\

Why does $G_\mathrm{eff}$ have such an impact on the fit to ACT+\textit{WMAP} data? As a single parameter, it influences the CMB power spectra in 3 ways: amplitude, phase, and tilt as described in Ref.~\cite{kreisch}. This can be seen by looking at the correlation between $G_\mathrm{eff}$ and the parameters $A_{\rm s}e^{-2\tau}$, $\theta_*$, and $n_{\rm s}$ (shown in \autoref{fig:posteriors_variants} in Appendix \ref{app:2D_post}). Changing the value of $G_\mathrm{eff}$ between the two modes of the distribution is, thus, able to accomplish the equivalent of adjusting 3 parameters. We note, however, that $G_\mathrm{eff}$'s impacts on the power spectra are coupled -- large $G_\mathrm{eff}$ causes a boost in amplitude which coincides with a shift towards small scales and a blue tilt (all compared to $\Lambda$CDM). With 3 parameters that can either increase or decrease in values, there are a total of 8 different combinations of changes to the power spectra. Remarkably, $G_\mathrm{eff}$ is able to capture the one combination\footnote{It captures only one combination rather than two because when $G_\mathrm{eff}$ is small, the physics is equivalent to $\Lambda \mathrm{CDM}$.} that is actually favored by the data (discussed more in \autoref{sec:Emode}).\\

{\bf The role of $\Omega_{\rm b}h^2$, $N_{\rm eff}$, and $n_{\rm s}$.}\label{sec:lessdamp} The three effects generated by $G_\mathrm{eff}$ are also able to compensate for other fluctuations seen in the ACT fits. As discussed in Ref.~\cite{actlike1}, ACT data alone prefer a $2.3-2.7\sigma$ low fluctuation of $\Omega_{\rm b}h^2$ and high fluctuation of $n_{\rm s}$ within $\Lambda \mathrm{CDM}$ compared to \textit{WMAP} and \textit{Planck}. The low $\Omega_{\rm b}h^2$ value damps the first acoustic peak and small angular scales, and also produces a slight phase shift towards larger scales (due to the reduced baryon loading of the primeval plasma). This small-scale damping is then compensated by the high $n_{\rm s}$ value, which tilts the spectra up. This movement occurs along a strong degeneracy line and is alleviated when large angular scales are added to ACT. Allowing also $N_\mathrm{eff}$ to vary for ACT data alone yields similarly low $\Omega_{\rm b}h^2$ values, but with an $n_{\rm s}$ value more similar to \textit{Planck} accompanied by a low $N_\mathrm{eff}$ centered at $\sim 2.3$, disfavoring the larger relativistic degrees of freedom $N_\mathrm{eff}=3.5$ at $4\sigma$, as discussed in Ref.~\cite{actlike1}. $N_\mathrm{eff}$ and $n_{\rm s}$ are highly degenerate due to large $N_\mathrm{eff}$ causing small-scale damping~\cite{Hou:2011ec} and higher $n_{\rm s}$ being able to undo this damping. 
While most of the $\Omega_{\rm b}h^2$-- $n_{\rm s}$ ACT fit can be shifted with \textit{WMAP} data, the low fluctuation in $N_\mathrm{eff}$ remains (ACT+\textit{WMAP} prefers $N_\mathrm{eff}$ values $\sim 2.3\sigma$ lower than the standard value) and leaves room for the neutrino strong mode. 

Strong neutrino self-interactions, i.e.,~large values of $G_{\rm eff}$, are able to undo the damping and phase shift from the low preferred $\Omega_{\rm b}h^2$ by boosting the spectra and shifting it towards small scales. This is done by exploiting the multi-parameter degeneracy between $G_{\rm eff}$, $A_{\rm s}$ and $n_{\rm s}$ described in Ref.~\cite{lancaster}, which results in lower values of both $A_{\rm s}$ and $n_{\rm s}$ as compared to $\Lambda$CDM. At the same time, strong neutrino self-interactions allow $N_{\rm eff}$ to be closer to its standard value due to absence of free-streaming phase and amplitude shift \cite{Bashinsky:2003tk} on the CMB in this case -- see \autoref{tab:bestfit_sinu_baseline} in Appendix~\ref{app:params}. 

Appendix~\ref{sec:helium} explores the additional degeneracy with the primordial Helium abundance.

\section{The Role of E-mode Polarization}\label{sec:Emode}

\begin{figure}[t]
\begin{center}
\includegraphics[width=0.4\textwidth]{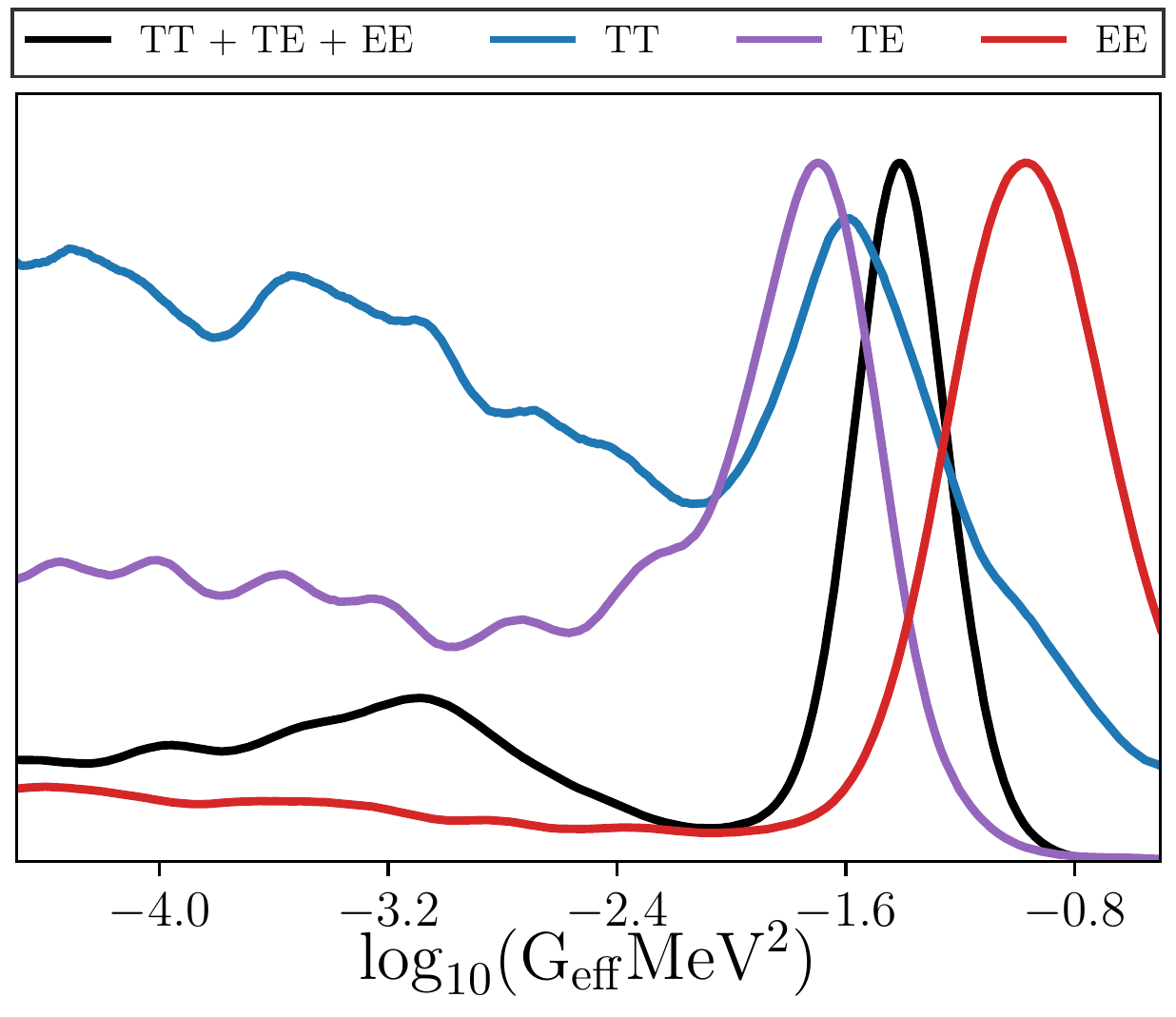}
\caption{ACT 1D posteriors for $G_{\rm eff}$ in the baseline model for the full TT+TE+EE data and for fits of separate spectra. We note again that we minimally smooth the posteriors, so low values of $G_{\rm eff}$ can appear unconverged.}
\label{fig:3components_act}
\end{center}
\end{figure}

ACT's preference for a delayed onset of neutrino free streaming is predominantly driven by its E-mode polarization spectrum. We show in \autoref{tab:chi2lcdm_sinu} the $\Delta \chi^2$ between SI$\nu$ (9 parameters) and $\Lambda \mathrm{CDM}$ (6 parameters) for different CMB data combinations. The largest $\Delta \chi^2$ in favor of the SI$\nu$ mode occurs for ACT's E-mode polarization data for ACT alone, with $\Delta \chi^2_{\rm ACT: EE}=-7.3$, and ACT+\textit{WMAP}, with $\Delta \chi^2_{\rm ACT: EE}=-6.0$. In \autoref{fig:3components_act} we show the $G_{\rm eff}$ posterior distribution for the ACT combined dataset and separate constraints from its TT, TE, and EE data.\footnote{These were obtained by running separate nested sampling runs with ACT TT data alone, TE data alone, and EE data alone.} While the TT and TE data show some preference for the SI$\nu$ mode, ACT's EE data dominates the preference for a delayed onset of neutrino free streaming as the MI$\nu$ mode nearly disappears in this case.

\begin{table}[t!]
  \caption{Comparison to $\Lambda\mathrm{CDM}$ for the strongly-interacting neutrino mode ($\Delta \chi^2 = \chi^2_{\mathrm{SI}\nu} - \chi^2_{\rm \Lambda CDM}$) for our baseline model. \label{tab:chi2lcdm_sinu}}
  \begin{ruledtabular}
  \begin{tabular}{ccccccc}
  Parameter & ACT & \textit{Planck} & ACT+\textit{WMAP} & ACT+\textit{Planck} \\
  \hline \\ [-2ex]
  $\Delta \chi^2_{\mathrm{ACT}}$ 	& $-10.0$	& --	& $-14.9$ & $-5.2$  \\
    $\Delta \chi^2_{\mathrm{ACT: TT}}$\footnote{Hereinafter, each $\Delta \chi^2_{\mathrm{ACT: X}}$ accounts once for the cross-correlation with the other components of ACT data such that the three $\Delta\chi^2_{\mathrm{ACT: X}}$ add up to $\Delta\chi^2_{\mathrm{ACT}}$ without overcounting.} 	& $-2.7$	& --	& $-1.9$ & $-1.1$   \\
  $\Delta \chi^2_{\mathrm{ACT: TE}}$ 	& $0.03$	& --	& $-7.0$ & $-4.3$   \\
  $\Delta \chi^2_{\mathrm{ACT: EE}}$ 	& $-7.3$	& --	& $-6.0$ & $0.2$   \\
  $\Delta \chi^2_{\mathrm{low\,}\ell}$ 	& --	& $3.1$	& -- & $4.9$ \\
  $\Delta \chi^2_{\mathrm{high\,}\ell}$ 	& --	& $0.4$	& -- & $2.0$  \\
  $\Delta \chi^2_{\mathrm{\textit{WMAP}}}$ 	& --	& --	& $0.8$ & -- \\
  $\Delta \chi^2_\mathrm{CMB\,Total}$ 	& $-10.0$	& $3.5$	& $-14.1$ & $1.7$ 
\\ [0.5ex] \hline \\[-2ex]
  $\Delta \chi^2_{\mathrm{prior}}$ 	& $0.09$	&  $-3.1$	& $0.4$ & $-3.4$  \\
  $\Delta \chi^2_\mathrm{Total}$ 	& $-10.0$	& $0.4$	& $-13.7$ & $-1.7$  \\
  $\Delta \mathrm{AIC}$ 	& $-4.0$	& $6.4$	& $-7.7$ & $4.3$
  \end{tabular}
\end{ruledtabular}
\end{table}

The addition of \textit{Planck} data neutralizes ACT's E-mode polarization preference. \textit{Planck} polarization data is signal dominated in windows below $\ell \approx 700$ (see Fig.~17 in Ref.~\cite{planck_diffuse}), and can therefore statistically overpower ACT's polarization preferences in this multipole range, which, as we will see below, plays a significant role in these constraints. {Given that \textit{WMAP} does not include EE data, the inclusion of \textit{WMAP} does not suppress ACT EE data's preference for SI$\nu$. Further, this combination also sees a preference for SI$\nu$ coming from the ACT TE data, with $\Delta \chi^2_{\rm ACT: TE}=-7.0$.}

\begin{figure*}[t]
\begin{center}
\includegraphics[width=0.9\textwidth]{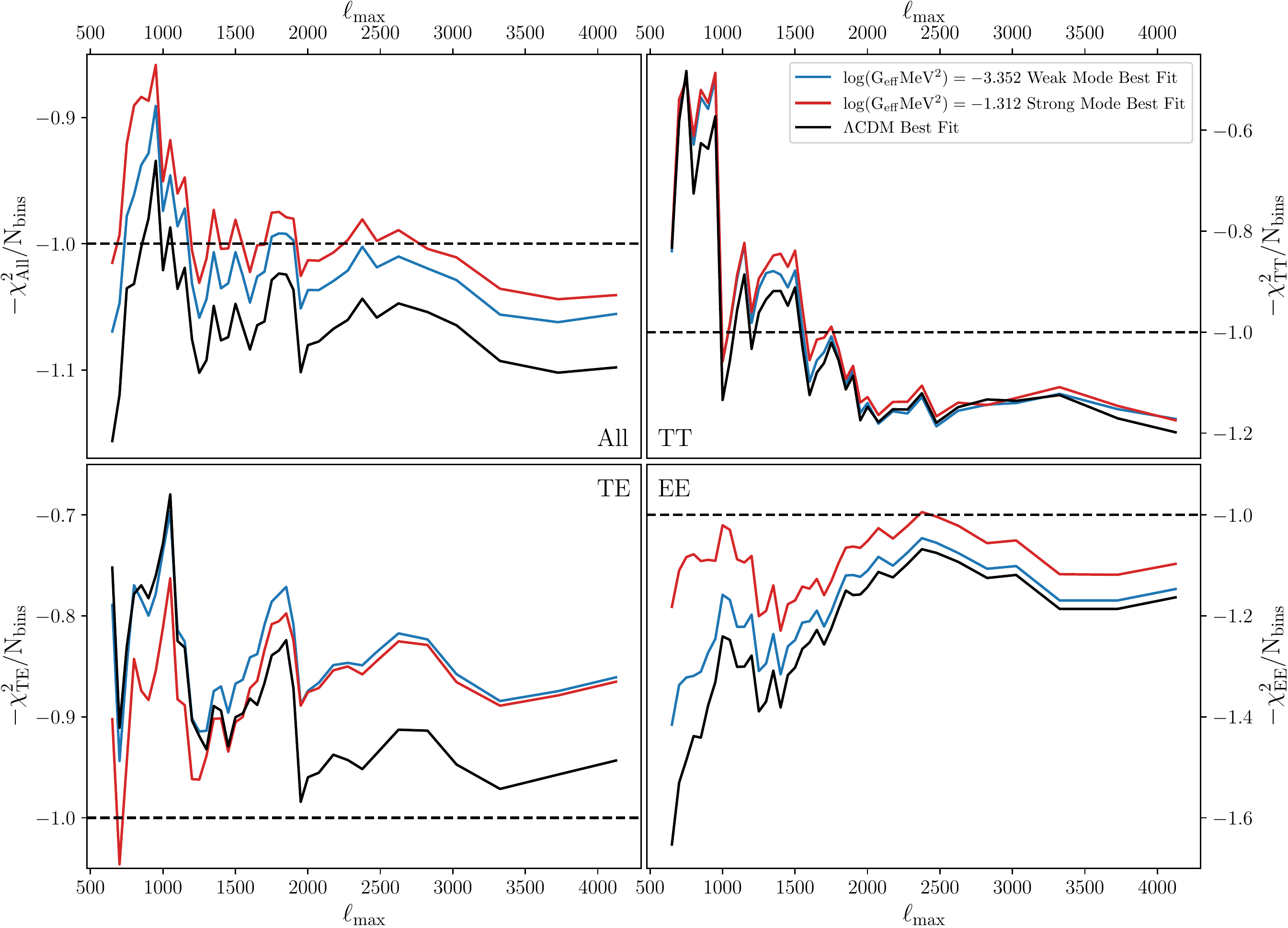}
\caption{$-\chi^2/{\rm N_{bins}}$ as a function of $\ell_{\rm max}$ where the former is the negative reduced $\chi^2$ and the latter the maximum $\ell$ used in the likelihood, shown for SI$\nu$ (in red), MI$\nu$ (in blue), and \LCDM\ (in black) best fits to ACT+\textit{WMAP} data. This is shown for the combined ACT data in top left, TT only in top right, TE only in bottom left, and EE only in bottom right. Overall, it is clear that for most choices of $\ell_{\rm max}$, SI$\nu$ has the largest $-\chi^2/{\rm N_{bins}}$ across all sectors of ACT data. The top left figure shows the greatest difference between SI$\nu$ and MI$\nu$ in the $700 \lesssim \ell \lesssim 1000$ range, which is echoed only by the bottom right, indicating the importance of E-mode polarization in that $\ell$ range in promoting the favorability of SI$\nu$.}
\label{fig:ACTWMAPchi2vlmax}
\end{center}
\end{figure*}

\begin{figure*}[t!]
\begin{center}
\includegraphics[width=\textwidth]{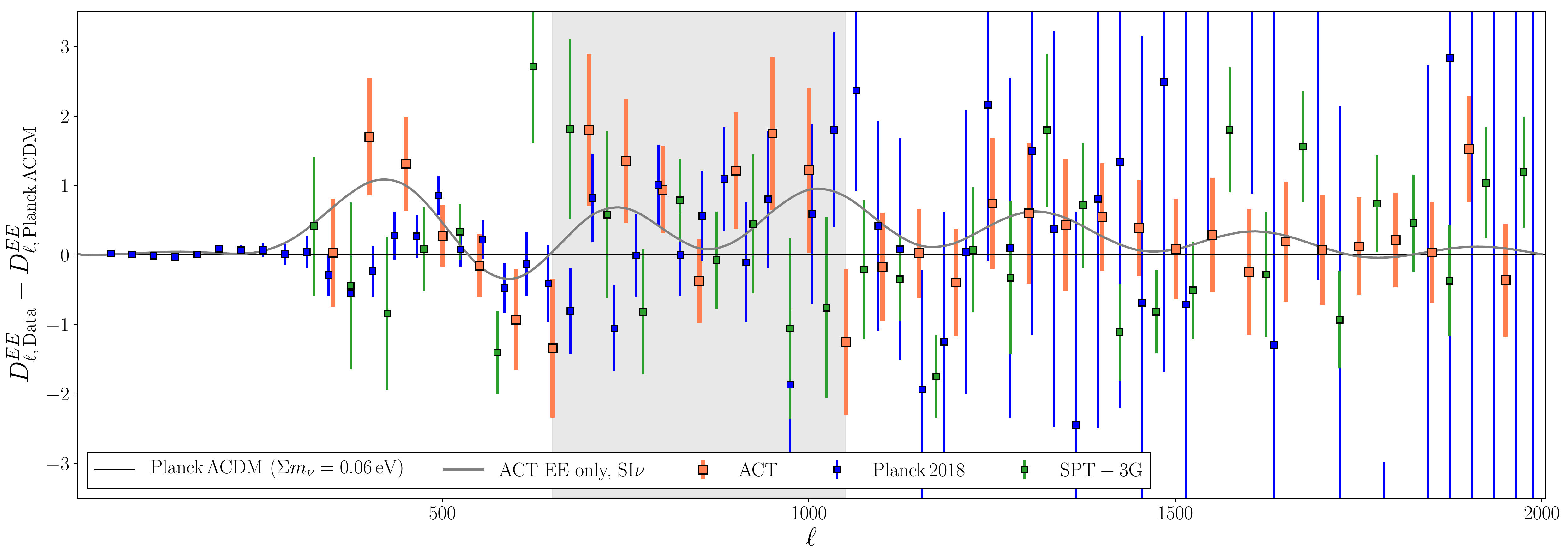}
\caption{Residuals of E-mode polarization data for ACT (orange), \textit{Planck} (blue), and SPT-3G (green) data relative to \textit{Planck}'s best-fit $\Lambda \mathrm{CDM}$ model. We overlay the best-fit for SI$\nu$ for just ACT E-mode polarization in grey (best-fit further described in \autoref{sec:Emode}), and the grey band highlights the multipoles that drive ACT's preference for SI$\nu$. Each bin spans 50 multipoles. Note the upwards deviation from the best fit in the $700\lesssim \ell \lesssim1000$ range of the ACT and SPT-3G data. The residuals of TT, TE, and EE power spectrum for the full $\ell$ range are presented in Appendix~\ref{app:CMBresidual}. }
\label{fig:data_diff}
\end{center}
\end{figure*}

Why does ACT's E-mode polarization prefer a delayed onset of neutrino free streaming while \textit{Planck}'s does not? To answer this question, we show the $-\chi^2$ per degree of freedom ($-\chi^2/{\rm d.o.f.}$) for the SI$\nu$ mode, MI$\nu$ mode, and $\Lambda \mathrm{CDM}$ as a function of the maximum $\ell$ used for the TT, TE, EE, and combined ACT data in \autoref{fig:ACTWMAPchi2vlmax}. We use the ACT+\textit{WMAP} best-fit models and we indicate with a dashed line a reference value of $-\chi^2/{\rm d.o.f.} = -1$. 

Overall, both the SI$\nu$ and MI$\nu$ modes provide a better fit than $\Lambda \mathrm{CDM}$ to the ACT E-mode polarization data (see Appendix \ref{app:fulltables} for tables comparing the MI$\nu$ mode to $\Lambda \mathrm{CDM}$). Both modes have larger $-\chi^2/{\rm d.o.f.}$ than $\Lambda \mathrm{CDM}$ in the bottom right panel of \autoref{fig:ACTWMAPchi2vlmax}, which focuses on the E-mode polarization. For the range $700 \lesssim \ell \lesssim 1000$ in this panel, as $\ell_{\rm max}$ increases the SI$\nu$ mode rapidly provides a better fit to the data than either the MI$\nu$ mode or $\Lambda \mathrm{CDM}$. The MI$\nu$ mode fits the TE cross-correlation data better than the SI$\nu$ mode in the same range, but the margin between the modes is twice as large with the EE autocorrelation. \emph{Ultimately the effect in the $700 \lesssim \ell \lesssim 1000$ range of the E-mode polarization drives the total likelihood in favor of the SI$\nu$ mode.} Beyond $\ell \sim 1000$, there is not significant information added by smaller scales to further differentiate the two interacting neutrino modes. Finally, angular scales corresponding to $\ell \gtrsim 1700$ do not incorporate additional information to further improve the fit of $\Lambda \mathrm{CDM}$ over the 2 modes. More specifically, we find that removing ACT data at $\ell > 2500$ does not significantly change the posterior distribution on $G_{\rm eff}$. By contrast, keeping only data at $\ell > 1000$ entirely removes the modest preference for $G_{\rm eff}\sim 10^{-1.6}$ MeV$^{-2}$ and leaves a nearly flat posterior, reinforcing the fact that the preference for the SI$\nu$ mode is driven by larger angular scales.

Our findings mirror those in Refs.~\cite{colinede,sterileint} showing that the most pertinent feature for the respective extensions to \LCDM\ is in the $\ell \lesssim 1000$ range of the EE spectrum of ACT data. 
This is also in agreement with the theoretical expectation provided in Ref.~\cite{Friedland:2007vv}, which concludes that CMB experiments are most sensitive to neutrino interactions at $\ell \lesssim 1000$.

Having identified this important range of angular scales in the E-mode polarization spectrum, we now study the residuals between the best-fit \textit{Planck} $\Lambda$CDM cosmological model and the three leading datasets of ACT, \textit{Planck}, and SPT-3G~\cite{SPT-3G:2021eoc} E-mode polarization data in this range. As shown in \autoref{fig:data_diff}, residuals for all experiments show a slight upward fluctuation in the range $700 \lesssim \ell \lesssim 1000$, with ACT showing the largest shift. A delayed onset of neutrino free streaming can better capture this upward fluctuation, driving the preference for the SI$\nu$ mode in the ACT data. Indeed, as described in Ref.~\cite{lancaster}, Fourier modes corresponding to this particular $\ell$ range enter the causal horizon close to the onset of neutrino free streaming for the SI$\nu$ mode, allowing them to be significantly influenced by the modified evolution of the gravitational potentials.  The figure also clarifies why the preference is reduced when including \textit{Planck} data: while consistent with both ACT and SPT-3G in this multipole range, \textit{Planck} does not possess such a strong E-mode upward fluctuation.

\section{CMB+Lensing+BAO results}\label{sec:BAO_H0}

We now expand our analysis beyond CMB-only results and as previously mentioned incorporate BAO and \textit{Planck} CMB lensing data. Overall we find that adding BAO measurements increases the significance of the SI$\nu$ compared to the MI$\nu$ mode.\footnote{A slight preference for the SI$\nu$ mode when including BAO and distance ladder $H_0$ measurements had been noted in Ref.~\cite{kreisch}, but only \textit{Planck} temperature data were included in their analysis. We see that ACT+\textit{WMAP} data when combined with BAO measurements, as well as when combined with local $H_0$ measurements from S$H_0$ES, have some preference for late neutrino free streaming. See Appendix~\ref{app:H0me} for an updated discussion on neutrino free streaming and local $H_0$ measurements in light of ACT data.} Such preference is not surprising for the ACT+\textit{WMAP} CMB dataset: the values of our baseline model's parameters most relevant to BAO (i.e.~$H_0$, $\Omega_{\rm m}$, and $N_{\rm eff}$) for ACT+\textit{WMAP} within the SI$\nu$ mode are closer to their concordance $\Lambda$CDM values than those within the MI$\nu$ mode and therefore BAO further consolidates this preference. 
We find that the SI$\nu$ mode has a $\Delta
\chi^2=-13.2$ compared to $\Lambda$CDM for ACT+\textit{WMAP} once BAO is included (Table \ref{tab:chi2lcdm_sinu_bao} in Appendix~\ref{app:fulltables}), roughly corresponding to a $2.9\sigma$ preference for a delayed onset of neutrino free streaming.
\LCDM\ continues to be preferred when adding BAO to ACT+\textit{Planck}. 

\section{Robustness Tests}\label{sec:syst}

We test the robustness of our results -- and in particular the contribution to them from ACT -- by examining different partitions of the ACT data independently, and revisiting some of the analysis assumptions.

\begin{itemize}
    \item \autoref{fig:3components_act} shows that different spectra (TT, TE, EE) in ACT give very consistent results.
    \item ACT data are further composed of `deep' maps, spanning 20-340 deg$^2$, and `wide' maps, spanning 210-1400 deg$^2$ (see Refs.~\cite{actlike1,actlike2}). Both sets contain TT, TE, and EE power spectra spanning the full $\ell$ ranges detailed in \autoref{sec:data}. After constraining the interaction with each set separately, we conclude that the two patches give consistent results but the wide data prefer the SI$\nu$ mode more than the deep data does. This weaker significance in the deep data is not unexpected, however, as the wide data has smaller errors at large scales, and therefore stronger constraining power, than the deep data at scales that are relevant for this model (See Appendix \ref{sec:ACTwidedeep} for further discussion). 
    \item The difference in mode preference between ACT and \textit{Planck} could also point to a different behaviour at large and small scales since the two experiments' constraining power peaks in different regimes~\cite{actlike1}. Even if ACT on its own does not yet allow for a high signal-to-noise comparison of the two regimes we run a few tests to check any effect on the results. We find that the \geff\ 1D posteriors when we vary the full 9 parameter space with all of ACT, or ACT small scales ($\ell>2500$) removed, or ACT large scales ($\ell<1000$) removed are all consistent. As mentioned before, for constraints of neutrino self-interaction with ACT data, the critical information is in the large scales and small scales hardly contribute to the demonstrated preference for SI$\nu$. This also supports the framework of a linear analysis.
    \item Finally, we note that multiplying or dividing the ACT TE power spectra by an artificial calibration factor of 1.05 to account for some unknown amplitude offset as in Ref.~\cite{actlike1} does not substantially change the results.  

\end{itemize}


\section{Conclusions}\label{sec:conc}

We have shown that ACT DR4 data, both alone and when combined with \textit{WMAP}, display a slight preference for a delayed onset of neutrino free streaming due to neutrino self interactions at the $2-3\sigma$ level. Like in previous analyses using CMB data, we find the parameter posteriors to be bimodal. When using ACT the strongly-interacting mode representing the delayed onset of neutrino free streaming has a higher likelihood than the \LCDM\ analogous, weakly-interacting mode. The preference reverts towards \LCDM\ when \textit{Planck} data is added to ACT, with the strongly-interacting mode suppressed and the bimodality moving more in favor of the \LCDM\ analogue model as seen in previous works. We determined that ACT data's preference is primarily driven by angular scale corresponding to $700\lesssim\ell\lesssim1000$ in the ACT E-mode polarization data. The impact of \textit{Planck} in the joint fit is mostly due to a wider range of cosmological information provided by \textit{Planck} and in particular in this EE range.

In summary, despite the high-resolution observations of the CMB temperature and polarization spectra we find the posterior to be persistently bimodal and the preference between the modes to be dependent on choice of model (${\rm G_{eff}}$ only or ${\rm G_{eff}}$+$\rm N_{eff}$+$\Sigma m_\nu$) and dataset combination. This indicates that we need more data to make any decisive judgements on this model, and motivates searches for unaccounted-for systematic effects in the $500\lesssim\ell\lesssim1000$ region in polarization that we highlighted (see also~Refs.~\cite{colinede, sterileint}). Alternative new-physics models could also be examined, including those where only a fraction of the neutrinos can self-interact (see e.g.,~Refs.~\cite{Das:2020xke,Mazumdar:2020ibx,Brinckmann:2020bcn}). 
Our results also call for detailed scrutiny of upcoming ACT polarization data, especially in the $\ell\lesssim1000$ region.

\acknowledgments
C.D.K.'s work is supported by the National Science Foundation (NSF) Graduate Research Fellowship under Grant DGE 1656466. EC acknowledges support from the European Research Council (ERC) under the European Union’s Horizon 2020 research and innovation programme (Grant agreement No. 849169). The work of F.-Y.~C.-R is supported by the NSF under grant AST-2008696. JD is supported by NSF grants 1814971 and 2108126. We would like to thank the UNM Center for Advanced Research Computing, supported in part by the NSF, for providing some of the research computing resources used in this work.  Part of this work was performed at the Aspen Center for Physics, which is supported by NSF grant PHY-1607611. Part of this work was
done at the Jet Propulsion Laboratory, California Institute of Technology, under a contract with the National Aeronautics and Space Administration. JCH acknowledges support from NSF grant AST-2108536.
ADH acknowledges support from the Sutton Family Chair in Science, Christianity and Cultures and from the Faculty of Arts and Science, University of Toronto. 
KM acknowledges support from the National Research Foundation of South Africa. Research at Perimeter Institute is supported in part by the Government of Canada through the Department of Innovation, Science and Industry Canada and by the Province of Ontario through the Ministry of Colleges and Universities.

Support for ACT was through the U.S.~National Science Foundation through awards AST-0408698, AST-0965625, and AST-1440226 for the ACT project, as well as awards PHY-0355328, PHY-0855887 and PHY-1214379. Funding was also provided by Princeton University, the University of Pennsylvania, and a Canada Foundation for Innovation (CFI) award to UBC.  ACT operates in the Parque Astron\'omico Atacama in northern Chile under the auspices of the Agencia Nacional de Investigaci\'on y Desarrollo (ANID).  The development of multichroic detectors and lenses was supported by NASA grants NNX13AE56G and NNX14AB58G.  Detector research at NIST was supported by the NIST Innovations in Measurement Science program. 

We acknowledge use of the {\tt matplotlib}~\cite{Hunter2007}, {\tt numpy}~\cite{2020Natur.585..357H}, {\tt GetDist}~\cite{Lewis:2019xzd}, and {\tt CosmoMC}~\cite{cosmomc} packages and use of the Boltzmann code {\tt CAMB}~\cite{CAMB}.

\begin{widetext}
\appendix
\section{Parameter Values}\label{app:params}
In this appendix, we list the credible intervals for the different model parameters used in our analyses. In \autoref{tab:bestfit_sinu_baseline}, we give the mean parameter values and their $1\sigma$ error bars for the SI$\nu$ mode for four data combinations. The corresponding MI$\nu$ values are given in \autoref{tab:bestfit_minu_baseline}. The parameter values within our simpler ${\rm G}_{\rm eff}$ one-parameter extension are given in \autoref{tab:bestfit_sinu_geff}. \autoref{tab:chi2lcdm_awext} shows the $\Delta \chi^2$ between \LCDM\ and its extension with $\Sigma m_\nu+N_{\rm eff}$ and (the strong mode of) only $G_{\rm eff}$.  For both models the SI$\nu$ mode prefers slightly higher values of $H_0$ and $\sigma_8$ than the MI$\nu$ mode given all the listed dataset combinations. Overall the estimated values of $H_0$ are lower than in \LCDM\ due to its anti-correlation with neutrino mass, which is held fixed in \LCDM. 

\begin{table*}[h!]
  \caption{Credible intervals ($68\%$) for the SI$\nu$ mode for our baseline model. \label{tab:bestfit_sinu_baseline}}
  \begin{ruledtabular}
  \begin{tabular}{ccccccc}
  Parameter & ACT & \textit{Planck} & ACT + \textit{WMAP} & ACT + \textit{Planck} \\
  \hline \\ [-2ex]
{$\Omega_{\rm b} h^2   $} & $0.02100\pm 0.00050        $& $0.02240\pm 0.00023        $ & $0.02188\pm 0.00035        $ & $0.02214^{+0.00024}_{-0.00022}$\\

{$\Omega_{\rm c} h^2   $} & $0.1164^{+0.0069}_{-0.0079}$& $0.1165^{+0.0025}_{-0.0036}$& $0.1195\pm 0.0053          $& $0.1171\pm 0.0031          $\\

{$100\theta_{\rm MC} $} & $1.04709\pm 0.00090        $ & $1.04564^{+0.00054}_{-0.00062}$ & $1.04757\pm 0.00077        $ & $1.04615\pm 0.00046        $\\

{$\tau           $} & $0.065\pm 0.014            $ & $0.0736^{+0.012}_{-0.0067} $ & $0.060^{+0.011}_{-0.012}   $ & $0.071^{+0.012}_{-0.010}   $\\

{$\sum m_\nu \,[{\rm eV}]  $} & $0.83^{+0.60}_{-0.28}      $& $0.106^{+0.029}_{-0.11}    $& $0.59\pm 0.30              $& $0.246^{+0.080}_{-0.24}    $\\

{$N_{\rm eff}        $} & $2.45^{+0.40}_{-0.55}      $ & $2.79^{+0.16}_{-0.21}      $ & $2.82^{+0.28}_{-0.33}      $ & $2.74^{+0.17}_{-0.20}      $\\

{$\rm{log}_{10}( \it{G}_{\rm{eff}}{\rm MeV}^2)$} & $-1.42^{+0.19}_{-0.16}     $ & $-1.73^{+0.13}_{-0.17}     $ & $-1.29\pm 0.11             $ & $-1.61^{+0.11}_{-0.077}    $\\

{${\rm{ln}}(10^{10} A_{\rm s})$} & $2.983\pm 0.032            $ & $3.015^{+0.025}_{-0.015}   $ & $2.976\pm 0.027            $ & $3.010^{+0.026}_{-0.021}   $\\

{$n_{\rm s}            $} & $0.913^{+0.027}_{-0.040}   $ & $0.9312^{+0.0079}_{-0.0070}$ & $0.921\pm 0.013            $ & $0.9267\pm 0.0078          $\\
\hline \\[-1.5ex]



$H_0\,[{\rm km/s/Mpc}]                       $ & $56.5^{+3.5}_{-7.1}        $& $66.7\pm 1.9               $& $62.3^{+3.3}_{-4.7}        $& $64.7^{+3.1}_{-2.1}        $\\

$\Omega_{\rm m}                  $ & $0.471^{+0.10}_{-0.089}    $ & $0.315^{+0.010}_{-0.017}   $ & $0.385^{+0.050}_{-0.059}   $ & $0.341^{+0.015}_{-0.039}   $\\

$\sigma_8                  $ & $0.670^{+0.042}_{-0.094}   $ & $0.817^{+0.028}_{-0.011}   $ & $0.704^{+0.046}_{-0.072}   $ & $0.785^{+0.055}_{-0.022}   $\\



$Y_{\rm P}                       $ & $0.2361^{+0.0066}_{-0.0077}$ & $0.2418^{+0.0024}_{-0.0030}$ & $0.2420\pm 0.0043          $ & $0.2410\pm 0.0027          $\\

$r_*\,[{\rm Mpc}]                       $ & $149.2\pm 4.7              $ & $146.7^{+2.2}_{-1.6}       $ & $145.9\pm 3.0              $ & $146.9\pm 1.9              $\\




  \end{tabular}
\end{ruledtabular}
\end{table*}
\begin{table*}[htb!]
  \caption{Credible intervals ($68\%$) for the MI$\nu$ mode for our baseline model. \label{tab:bestfit_minu_baseline}}
  \begin{ruledtabular}
  \begin{tabular}{ccccccc}
  Parameter & ACT & \textit{Planck} & ACT + \textit{WMAP} & ACT + \textit{Planck} \\
  \hline \\ [-2ex]
{$\Omega_{\rm b} h^2   $} & $0.02064^{+0.00040}_{-0.00045}$& $0.02229\pm 0.00023        $& $0.02153^{+0.00030}_{-0.00035}$& $0.02200\pm 0.00021        $\\

{$\Omega_{\rm c} h^2   $} & $0.1092^{+0.0053}_{-0.0063}$& $0.1186\pm 0.0029          $& $0.1119^{+0.0044}_{-0.0051}$& $0.1155\pm 0.0025          $\\

{$100\theta_{\rm MC} $} & $1.04341^{+0.00099}_{-0.0011}$& $1.04105\pm 0.00044        $& $1.04255\pm 0.00082        $& $1.04153\pm 0.00039        $\\

{$\tau           $} & $0.065\pm 0.013            $& $0.0749^{+0.014}_{-0.0099} $& $0.060^{+0.011}_{-0.012}   $& $0.071^{+0.013}_{-0.010}   $\\

{$\sum m_\nu\,[{\rm eV}] $} & $< 0.771                   $ & $0.112^{+0.028}_{-0.11}    $ & $0.70^{+0.30}_{-0.26}      $ & $0.171^{+0.050}_{-0.17}    $\\

{$N_{\rm eff}        $} & $2.03^{+0.28}_{-0.42}      $ & $2.95\pm 0.19              $ & $2.37^{+0.25}_{-0.28}      $ & $2.71\pm 0.16              $\\

{$\rm{log}_{10}( \it{G}_{\rm{eff}}{\rm MeV}^2)$} & $-3.13^{+0.45}_{-0.62}     $ & $-5.89^{+0.99}_{-1.8}      $ & $-4.9^{+2.2}_{-2.5}        $ & $-5.7\pm 1.3               $\\

{${\rm{ln}}(10^{10} A_{\rm s})$} & $3.019\pm 0.035            $ & $3.081^{+0.029}_{-0.022}   $ & $3.029\pm 0.028            $ & $3.069^{+0.027}_{-0.023}   $\\

{$n_{\rm s}            $} & $0.905^{+0.024}_{-0.038}   $& $0.9616\pm 0.0087          $& $0.933\pm 0.014            $& $0.9523\pm 0.0082          $\\
\hline \\[-1.5ex]




$H_0 \,[{\rm km/s/Mpc}]                      $ & $52.0^{+2.4}_{-5.2}        $ & $66.3^{+2.0}_{-1.5}        $ & $56.9^{+2.4}_{-3.8}        $ & $63.9^{+2.4}_{-1.5}        $\\

$\Omega_{\rm m}                  $ & $0.529^{+0.10}_{-0.075}    $ & $0.323^{+0.010}_{-0.020}   $ & $0.440\pm 0.055            $ & $0.342^{+0.012}_{-0.029}   $\\

$\sigma_8                  $ & $0.631^{+0.030}_{-0.076}   $ & $0.813^{+0.029}_{-0.014}   $ & $0.665^{+0.036}_{-0.062}   $ & $0.790^{+0.041}_{-0.017}   $\\



$Y_{\rm P}                       $ & $0.2297^{+0.0050}_{-0.0064}$& $0.2441\pm 0.0026          $& $0.2354\pm 0.0041          $& $0.2405\pm 0.0023          $\\

$r_*\,[{\rm Mpc}]                       $ & $153.7^{+4.1}_{-3.6}       $ & $145.3\pm 1.8              $ & $150.6\pm 3.0              $ & $147.7\pm 1.6              $\\




  \end{tabular}
\end{ruledtabular}
\end{table*}
\begin{table*}[ht!]
  \caption{Credible intervals ($68\%$) for the SI$\nu$ mode for our simple $\mathrm{G}_{\rm eff}$ model. \label{tab:bestfit_sinu_geff}}
  \begin{ruledtabular}
  \begin{tabular}{cccccc}
  Parameter & ACT & ACT + \textit{WMAP} & ACT + \textit{Planck} \\
  \hline \\ [-2ex]
{$\Omega_{\rm b} h^2   $} & $0.02155\pm 0.00031        $& $0.02222\pm 0.00019        $& $0.02246\pm 0.00014        $\\

{$\Omega_{\rm c} h^2   $} & $0.1209\pm 0.0041          $ & $0.1211\pm 0.0026          $ & $0.1214\pm 0.0013          $\\

{$100\theta_{\rm MC} $} & $1.0462^{+0.0026}_{-0.0042}$& $1.04793\pm 0.00071        $& $1.04639^{+0.00045}_{-0.00038}$\\

{$\tau           $} & $0.063\pm 0.013            $& $0.059^{+0.010}_{-0.012}   $& $0.0695^{+0.012}_{-0.0084} $\\

{$\rm{log}_{10}( \it{G}_{\rm{eff}}{\rm MeV}^2)$} & $-2.3^{+1.5}_{-1.1}        $& $-1.277\pm 0.090           $& $-1.552\pm 0.081           $\\

{${\rm{ln}}(10^{10} A_{\rm s})$} & $3.000^{+0.031}_{-0.051}   $& $2.975\pm 0.021            $& $3.011^{+0.023}_{-0.018}   $\\

{$n_{\rm s}            $} & $0.969^{+0.018}_{-0.031}   $& $0.9334\pm 0.0060          $& $0.9375^{+0.0050}_{-0.0043}$\\

\hline \\[-1.5ex]




$H_0\,[{\rm km/s/Mpc}]                       $ & $68.2\pm 1.6               $ & $69.3\pm 1.1               $ & $68.84\pm 0.56             $\\

$\Omega_{\rm m}                  $ & $0.308^{+0.020}_{-0.024}   $ & $0.300\pm 0.015            $ & $0.3051\pm 0.0075          $\\

$\sigma_8                  $ & $0.832\pm 0.017            $ & $0.813\pm 0.013            $ & $0.8361\pm 0.0092          $\\








  \end{tabular}
\end{ruledtabular}
\end{table*}

\newpage

 \begin{table}[t!]
 \begin{minipage}[b]{0.5\linewidth}
  \caption{ACT + \textit{WMAP} extension comparisons to $\Lambda\mathrm{CDM}$  ($\Delta \chi^2 = \chi^2_{\rm ext} - \chi^2_{\rm \Lambda CDM}$) \label{tab:chi2lcdm_awext}}
  \begin{ruledtabular}
  \begin{tabular}{ccc}
  Parameter & $\sum m_{\nu}$ + $N_{\rm eff}$ & $G_{\rm eff}$ SI$\nu$ \\
  \hline \\ [-2ex]
  $\Delta \chi^2_{\mathrm{ACT}}$ 	& $-9.0$	& $-12.1$	 \\
  $\Delta \chi^2_{\mathrm{ACT: TT}}$	& $-1.4$	& $-3.2$ \\
  $\Delta \chi^2_{\mathrm{ACT: TE}}$ 	& $-6.2$	& $-2.8$ \\
  $\Delta \chi^2_{\mathrm{ACT: EE}}$ 	& $-1.4$	& $-6.2$	\\
  $\Delta \chi^2_{\mathrm{\textit{WMAP}}}$ 	& $0.5$	& $-0.3$	 \\
  $\Delta \chi^2_\mathrm{CMB\,Total}$ 	& $-8.6$	& $-12.4$
\\ [0.5ex] \hline \\[-2ex]
  $\Delta \chi^2_{\mathrm{prior}}$ 	& $0.1$	&  $0.3$ \\
  $\Delta \chi^2_\mathrm{Total}$ 	& $-8.4$	& $-12.1$ \\
  $\Delta \mathrm{AIC}$ 	& $-4.4$	& $-10.1$
  \end{tabular}
\end{ruledtabular}
\end{minipage}
\end{table}

\section{Mode Comparison}\label{app:fulltables}

\begin{table*}[t]
  \caption{Mode comparison. $\mathcal{B}_{\mathrm{SI}\nu}$ is the Bayes factor between the SI$\nu$ and the MI$\nu$ modes, $\mathcal{R}_{\mathrm{SI}\nu}$ is the maximum likelihood ratio, and $\Delta \chi^2 = \chi^2_{\mathrm{SI}\nu} - \chi^2_{\mathrm{MI}\nu}$.  \label{tab:ratios_full}}
  \begin{ruledtabular}
  \begin{tabular}{ccccc}
  Parameter & ACT & \textit{Planck} & ACT + \textit{WMAP} & ACT + \textit{Planck} \\ 
  \hline \\ [-2ex]
  $\mathcal{B}_{\mathrm{SI}\nu}$ 	& $1.5 \pm  0.2$	& $0.01  \pm 0.01$	& $2.8  \pm  0.6$ & $0.05  \pm 0.02$  \\
  $\mathcal{R}_{\mathrm{SI}\nu}$ 	& $2.6$	& $0.4$	& $6.5$ & $1.1$  \\ [0.5ex]
  \hline \\[-2ex]
  $\Delta \chi^2_{\mathrm{ACT}}$ 	& $-1.9$	& --	& $-3.9$ & $-1.6$ \\
  $\Delta \chi^2_{\mathrm{ACT: TT}}$ 	& $-0.06$	& --	& $0.2$ & $0.4$ \\
  $\Delta \chi^2_{\mathrm{ACT: TE}}$ 	& $-0.6$	& --	& $0.4$ & $-2.2$ \\
  $\Delta \chi^2_{\mathrm{ACT: EE}}$ 	& $-1.3$	& --	& $-4.5$ & $0.3$\\
  $\Delta \chi^2_{\mathrm{low\,}\ell}$ 	& --	& $2.8$	& -- & $3.9$ \\
  $\Delta \chi^2_{\mathrm{high\,}\ell}$ 	& --	& $1.0$	& -- & $3.2$ \\
  $\Delta \chi^2_{\mathrm{\textit{WMAP}}}$ 	& --	& --	& $0.1$ & --  \\
$\Delta \chi^2_\mathrm{CMB\,Total}$ 	& $-1.9$	& $3.8$	& $-3.8$ & $5.5$ 
\\ [0.5ex] \hline \\[-2ex]
  $\Delta \chi^2_{\mathrm{prior}}$ 	& $0.03$	&  $-2.0$	& $8.6\times 10^{-3}$ & $-5.7$ \\
  
  $\Delta \chi^2_\mathrm{Total}$ 	& $-1.9$	& $1.8$	& $-3.7$ & $-0.2$ 
  \end{tabular}
\end{ruledtabular}
\end{table*}
\begin{table}[h!]{}
\begin{minipage}{0.49\linewidth}
  \caption{Mode comparison for data combinations containing BAO measurements. $\mathcal{B}_{\mathrm{SI}\nu}$ is the Bayes factor between the SI$\nu$ and the MI$\nu$ modes, $\mathcal{R}_{\mathrm{SI}\nu}$ is the maximum likelihood ratio, and $\Delta \chi^2 = \chi^2_{\mathrm{SI}\nu} - \chi^2_{\mathrm{MI}\nu}$.  \label{tab:ratios_full_bao}}
  \begin{ruledtabular}
  \begin{tabular}{ccc}
  Parameter & ACT + \textit{WMAP} + BAO & ACT + \textit{Planck} + BAO \\ 
  \hline \\ [-2ex]
  $\mathcal{B}_{\mathrm{SI}\nu}$ 	&  $17.2  \pm  4.7$ & $0.1 \pm 0.04$ \\
  $\mathcal{R}_{\mathrm{SI}\nu}$ 	&  $35.8$ & $0.5$ \\ [0.5ex]
  \hline \\[-2ex]
  $\Delta \chi^2_{\mathrm{ACT}}$ 	&  $-5.8$ & $1.5$ \\
  $\Delta \chi^2_{\mathrm{ACT: TT}}$ 	&  $-0.5$  & $-0.7$ \\
  $\Delta \chi^2_{\mathrm{ACT: TE}}$ 	& $0.2$ & $1.5$ \\
  $\Delta \chi^2_{\mathrm{ACT: EE}}$ 	&  $-5.5$ & $0.8$ \\
  $\Delta \chi^2_{\mathrm{low\,}\ell}$ 	&  -- & $3.8$ \\
  $\Delta \chi^2_{\mathrm{high\,}\ell}$ 	&  -- & $-1.3$\\
  $\Delta \chi^2_{\mathrm{Lensing}}$ 	&  -- & $0.7$\\
  $\Delta \chi^2_{\mathrm{\textit{WMAP}}}$ 	&  $-1.1$ & -- \\
  $\Delta \chi^2_{\mathrm{BAO}}$ 	&  $-0.3$ & $-0.7$ \\
$\Delta \chi^2_\mathrm{CMB\,Total}$ 	&  $-6.9$ & $4.7$
\\ [0.5ex] \hline \\[-2ex]
  $\Delta \chi^2_{\mathrm{prior}}$ 	& $0.07$ & $-2.7$\\
  
  $\Delta \chi^2_\mathrm{Total}$ 	&  $-7.2$ & $1.3$
  \end{tabular}
\end{ruledtabular}
\end{minipage}
\begin{minipage}{0.49\linewidth}
  \caption{Comparison to $\Lambda\mathrm{CDM}$ for the Strongly Interacting Neutrino Mode for data combinations containing BAO measurements ($\Delta \chi^2 = \chi^2_{\mathrm{SI}\nu} - \chi^2_{\rm \Lambda CDM}$) \label{tab:chi2lcdm_sinu_bao}}
  \begin{ruledtabular}
  \begin{tabular}{ccc}
  Parameter &  ACT + \textit{WMAP} + BAO & ACT + \textit{Planck} + BAO\\
  \hline \\ [-2ex]
  $\Delta \chi^2_{\mathrm{ACT}}$ 	&  $-13.0 $ & $-2.7$ \\
  $\Delta \chi^2_{\mathrm{ACT: TT}}$ 	& $-2.9$ & $-2.2$ \\
  $\Delta \chi^2_{\mathrm{ACT: TE}}$ 	&  $-4.1$ & $-1.0$ \\
  $\Delta \chi^2_{\mathrm{ACT: EE}}$ 	& $-6.0$ & $0.5$ \\
  $\Delta \chi^2_{\mathrm{low\,}\ell}$ 	&  -- & $5.6$\\
  $\Delta \chi^2_{\mathrm{high\,}\ell}$ 	&  -- & $-3.2$\\
  $\Delta \chi^2_{\mathrm{Lensing}}$ 	&  -- & $0.4$\\
  $\Delta \chi^2_{\mathrm{\textit{WMAP}}}$ 	&  $0.0$ & -- \\
  $\Delta \chi^2_{\mathrm{BAO}}$ 	&  $-0.9$ & $-0.3$\\
  $\Delta \chi^2_\mathrm{CMB\,Total}$ 	& $-13.0$ & $0.1$
\\ [0.5ex] \hline \\[-2ex]
  $\Delta \chi^2_{\mathrm{prior}}$ 	&  $0.7$ &$-2.0$ \\
  $\Delta \chi^2_\mathrm{Total}$ 	&  $-13.2$ & $-2.1$\\
  $\Delta \mathrm{AIC}$ 	& $-7.2$ &$3.9$
  \end{tabular}
\end{ruledtabular}
\vspace{0.6em}
\end{minipage}

\end{table}
In this appendix we supplement the comparison of the $\chi^2$ values of SI$\nu$, MI$\nu$, and \LCDM\ best-fit models with respect to different datasets and their components. \autoref{tab:ratios_full} takes the SI$\nu$ mode and MI$\nu$ best fits to ACT, \textit{Planck}, ACT+\textit{WMAP}, and ACT+Planck datasets, and shows the difference in $\chi^2$ between the two best fits for the components of the datasets. \autoref{tab:ratios_full_bao} does the same with ACT+\textit{WMAP}+BAO and ACT+Planck+BAO as the datasets. \autoref{tab:chi2lcdm_minu} takes the MI$\nu$ and \LCDM\ best fits to ACT, \textit{Planck}, ACT+\textit{WMAP}, and ACT+Planck datasets, and shows the difference in $\chi^2$ between the two best fits for the components of the datasets. \autoref{tab:chi2lcdm_sinu_bao} does the same with ACT+\textit{WMAP}+BAO and ACT+Planck+BAO as the datasets for the SI$\nu$ mode. Note that though MI$\nu$ provides a better fit to the data than \LCDM, it does not fit better than SI$\nu$. 
 \begin{table}[h!]
  \caption{Comparison to $\Lambda\mathrm{CDM}$ for the Moderately Interacting Neutrino Mode  ($\Delta \chi^2 = \chi^2_{\mathrm{MI}\nu} - \chi^2_{\rm \Lambda CDM}$) \label{tab:chi2lcdm_minu}}
  \begin{ruledtabular}
  \begin{tabular}{ccccc}
  Parameter & ACT & \textit{Planck} & ACT + \textit{WMAP} & ACT + \textit{Planck} \\
  \hline \\ [-2ex]
  $\Delta \chi^2_{\mathrm{ACT}}$ 	& $-8.1$	& --	& $-11.0$ & $-3.6$  \\
  $\Delta \chi^2_{\mathrm{ACT: TT}}$ 	& $-2.7$	& --	& $-2.1$ & $-1.5$  \\
  $\Delta \chi^2_{\mathrm{ACT: TE}}$ 	& $0.6$	& --	& $-7.4$ & $-2.1$  \\
  $\Delta \chi^2_{\mathrm{ACT: EE}}$ 	& $-6.0$	& --	& $-1.5$ & $-0.03$   \\
  $\Delta \chi^2_{\mathrm{low\,}\ell}$ 	& --	& $0.3$	& -- & $1.1$ \\
  $\Delta \chi^2_{\mathrm{high\,}\ell}$ 	& --	& $-0.6$	& -- & $-1.2$ \\
  $\Delta \chi^2_{\mathrm{\textit{WMAP}}}$ 	& --	& --	& $0.7$ & --  \\
  $\Delta \chi^2_\mathrm{CMB\,Total}$ 	& $-8.1$	& $-0.3$	& $-10.3$ & $-3.8$ 
\\ [0.5ex] \hline \\[-2ex]
  $\Delta \chi^2_{\mathrm{prior}}$ 	& $-7.5\times 10^{-4}$	&  $-1.1$	& $0.4$ & $2.3$ \\
  $\Delta \chi^2_\mathrm{Total}$ 	& $-8.1$	& $-1.4$	& $-10.0$ & $-1.5$ \\
  $\Delta \mathrm{AIC}$ 	& $-2.1$	& $4.6$	& $-4.0$ & $4.5$ 
  \end{tabular}
\end{ruledtabular}
\end{table}

\section{Complete Posterior Plots}\label{app:2D_post}

We present the triangle plot showing \loggeff, $N_{\rm eff}$, $H_0$, $\sigma_8$, $100\theta_*$, $10^9 A_se^{-2\tau}$, $n_s$, and $\Omega_b h^2$ in \autoref{fig:posteriors_variants}. The posteriors were generated by varying the parameter spaces indicated in the legend with respect to ACT and \textit{WMAP}. Note the correlation  between the value of \loggeff\ in each island, and the values of $100\theta_*$, $10^9 A_se^{-2\tau}$, and $n_s$, that is most clearly demonstrated by the red contours (\LCDM\ augmented by \loggeff\ only). This indicates that \loggeff\ simultaneously impacts phase, amplitude, and tilt respectively. 

The $\rm{G_{eff}}$ model does not change the posteriors of $H_0$ and $\sigma_8$ significantly from the \LCDM\ result but the other extensions do. The similarity of the $N_{\rm eff}+\sum m_\nu$ contours and the baseline model contours indicate that the shift to lower $H_0$ is largely due to the variation in  $N_{\rm eff}$ and $\sum m_\nu$. This is in contrast to the results of Ref.~\cite{colinede} where fitting the EDE model to ACT data resulted in a shift to higher values of $H_0$. Though the new physics introduced here and in Ref.~\cite{colinede} both appear compatible with ACT data and capture similar features, their relationships with the data are not identical.

\begin{figure}[ht!]
\begin{center}
\includegraphics[width=\textwidth]{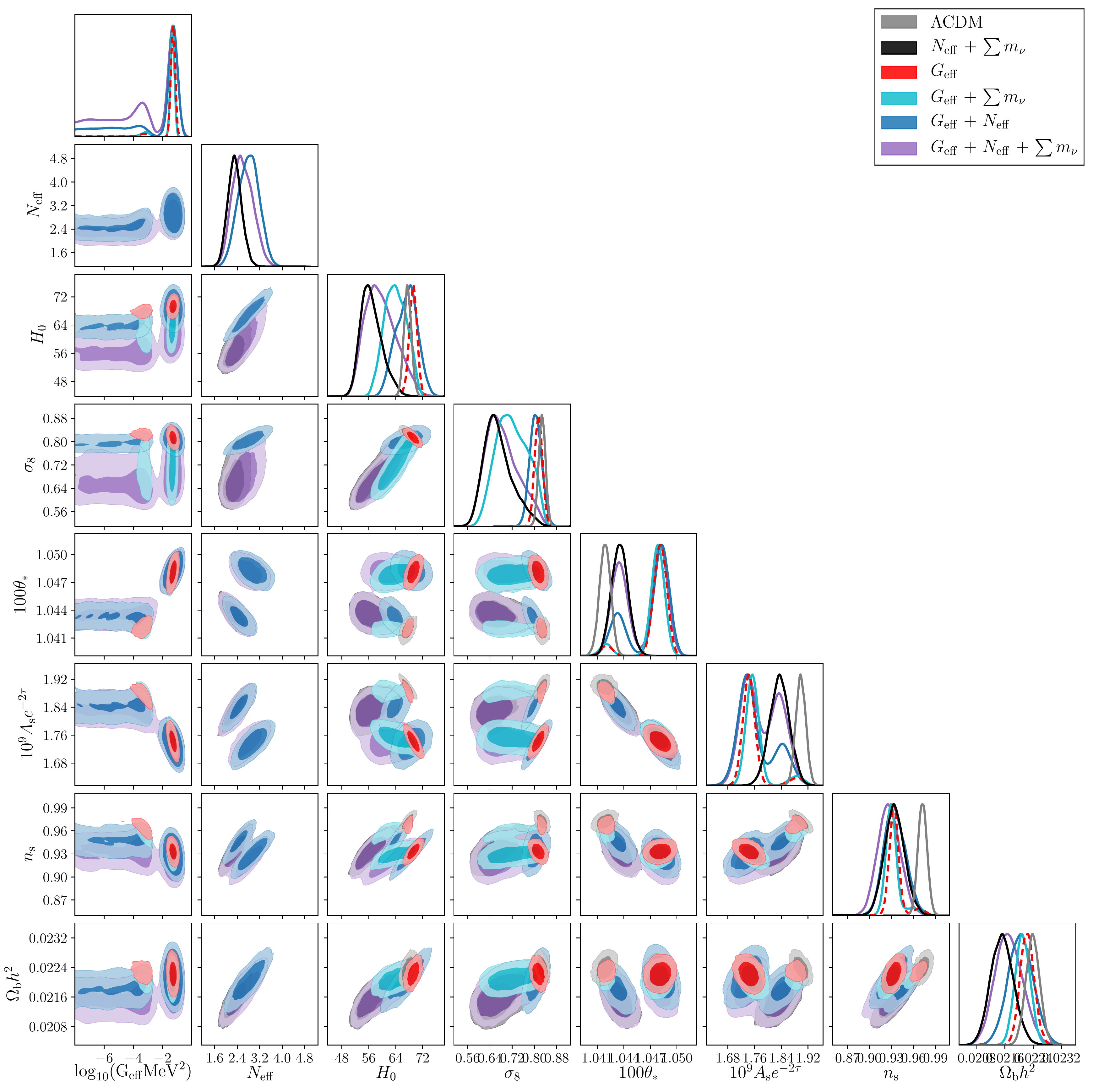}
\caption{Posterior distributions for \loggeff, $N_{\rm eff}$, $H_0$, $\sigma_8$, $100\theta_*$, $10^9 A_se^{-2\tau}$, $n_s$, and $\Omega_b h^2$ given ACT+\textit{WMAP} data and various extensions of \LCDM. The degeneracy between \loggeff\ and $100\theta_*$, $10^9 A_se^{-2\tau}$, and $n_s$ (in the direction between the peaks of the two modes) indicates \loggeff's ability to impact phase, amplitude, and tilt simultaneously}
\label{fig:posteriors_variants}
\end{center}
\end{figure}

\section{CMB Power Spectrum Residuals}\label{app:CMBresidual}

We present CMB power spectra from data and from theory predictions as residuals to the \LCDM\ best fit predictions. \autoref{fig:TT} shows the TT plots, \autoref{fig:TE} the TE plots, and \autoref{fig:EE} the EE plots. The left half of each Figure shows ACT data points and theory predictions from best fits to datasets including ACT. The right half shows \textit{Planck} data points and theory predictions from best fits to datasets including \textit{Planck}. The top half of each Figure shows SI$\nu$ best fits, and the bottom half MI$\nu$ best fits.

\begin{figure*}
\begin{center}
\includegraphics[width=0.48\textwidth]{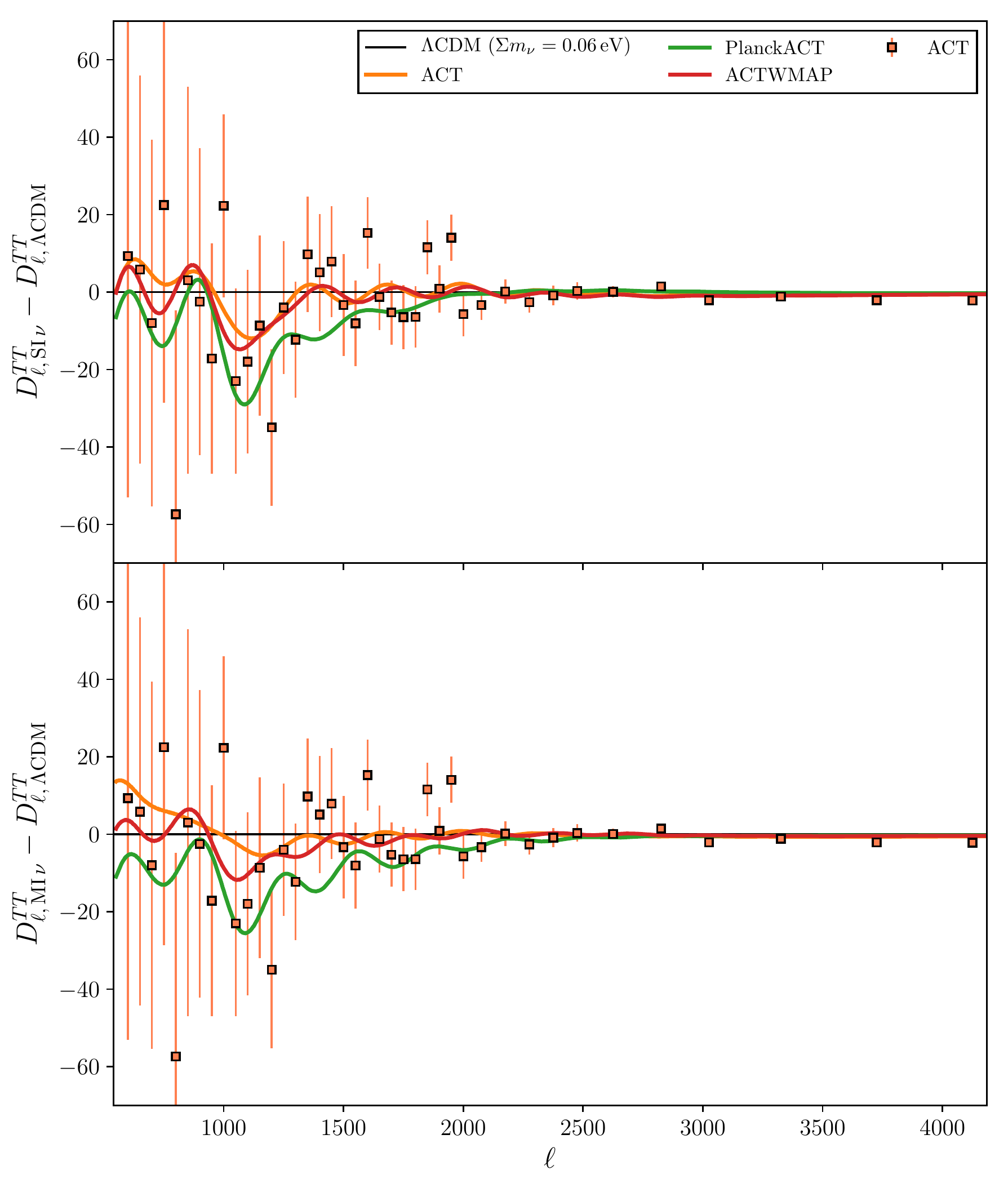}
\includegraphics[width=0.48\textwidth]{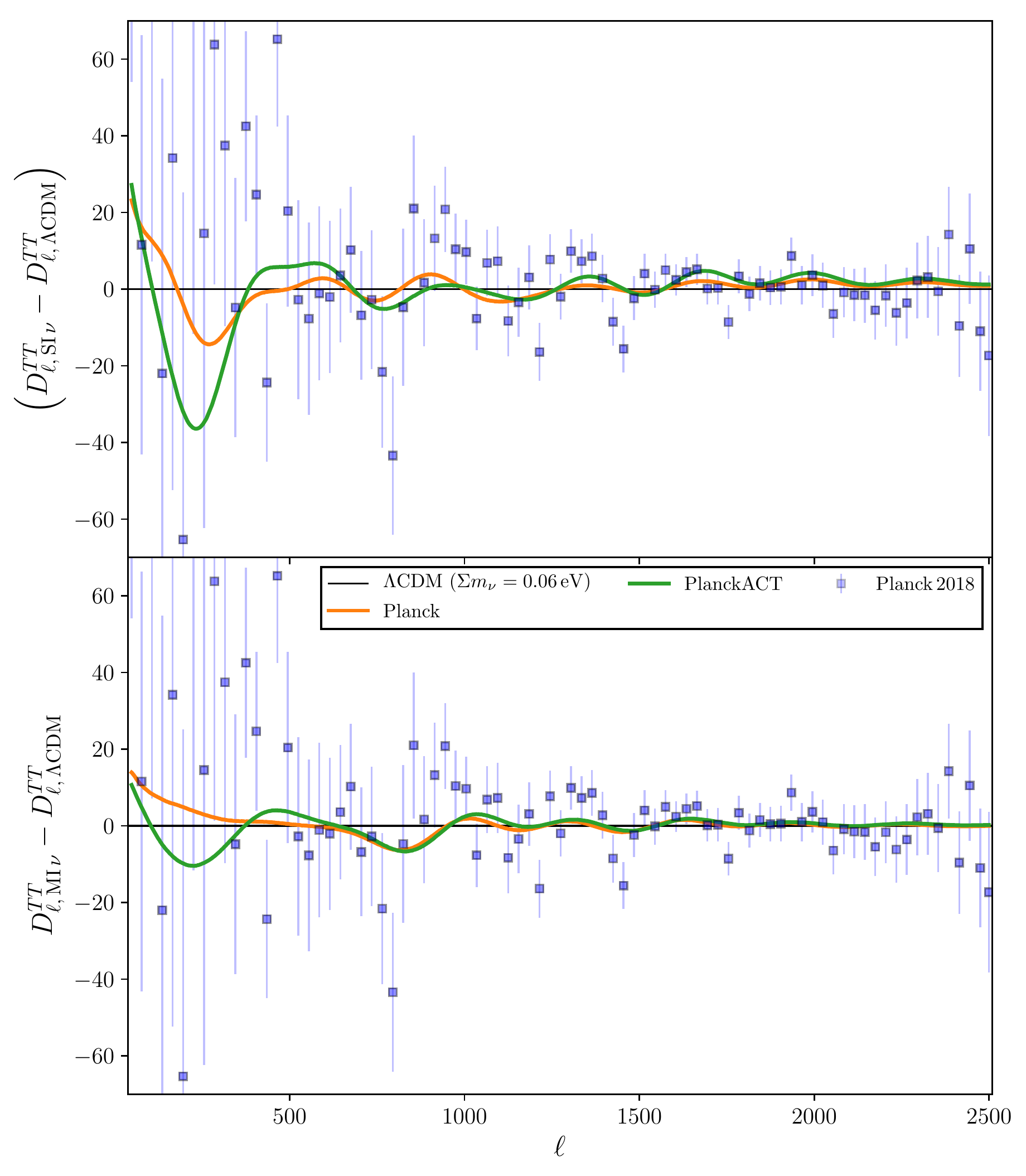}
\caption{Residuals of best-fit TT power spectra relative to $\Lambda \mathrm{CDM}$ (ACT best-fit $\Lambda \mathrm{CDM}$ for the left panel, \textit{Planck} best-fit $\Lambda \mathrm{CDM}$ for the right panel). The upper panels are SI$\nu$ best fits and the lower panels are MI$\nu$ best fits. }
\label{fig:TT}
\end{center}
\end{figure*}

\begin{figure*}
\begin{center}
\includegraphics[width=0.48\textwidth]{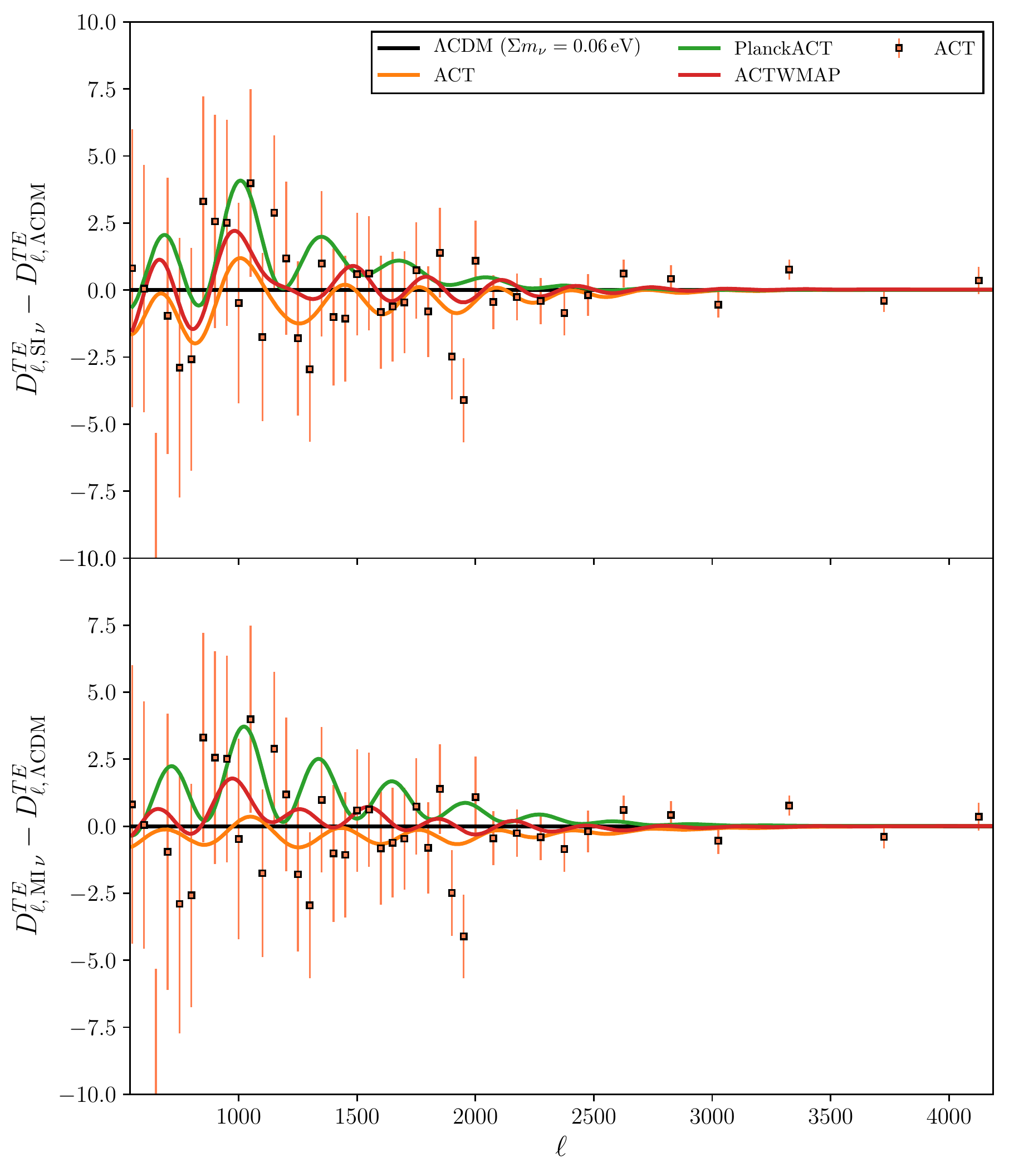}
\includegraphics[width=0.48\textwidth]{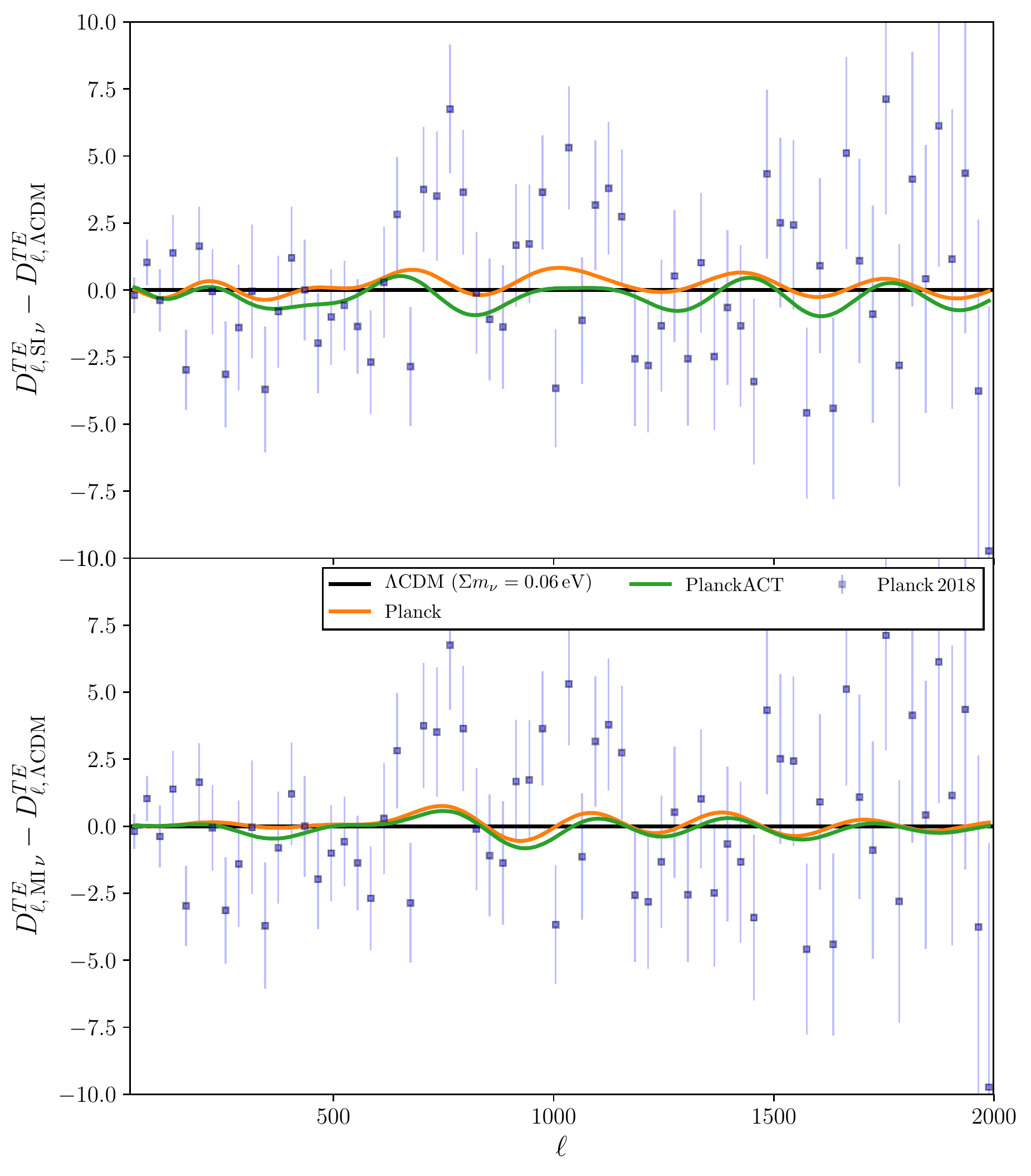}
\caption{Residuals of best-fit TE power spectra relative to $\Lambda \mathrm{CDM}$ (ACT best-fit $\Lambda \mathrm{CDM}$ for the left panel, \textit{Planck} best-fit $\Lambda \mathrm{CDM}$ for the right panel). The upper panels are SI$\nu$ best fits and the lower panels are MI$\nu$ best fits.}
\label{fig:TE}
\end{center}
\end{figure*}

\begin{figure*}
\begin{center}
\includegraphics[width=0.48\textwidth]{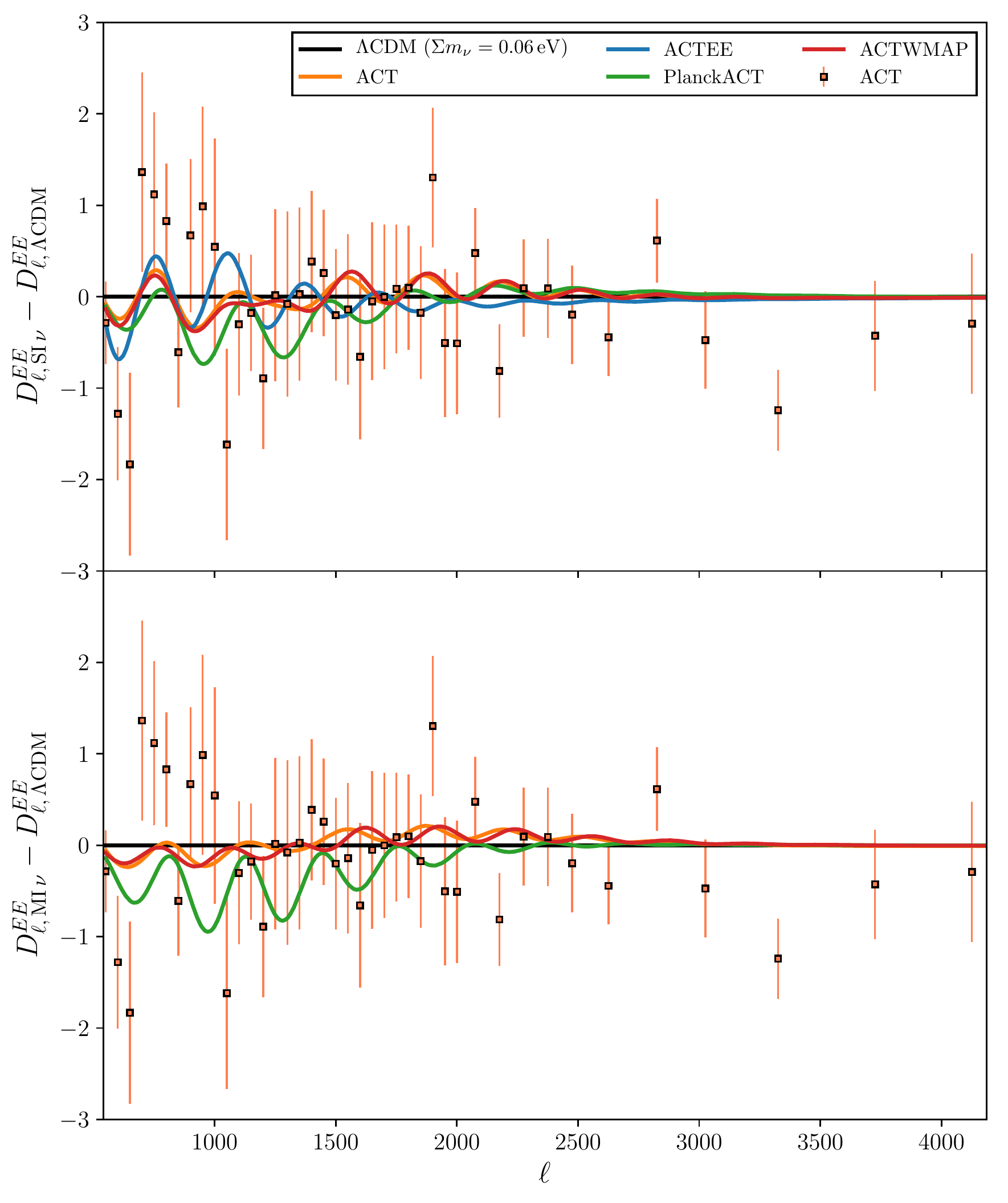}
\includegraphics[width=0.48\textwidth]{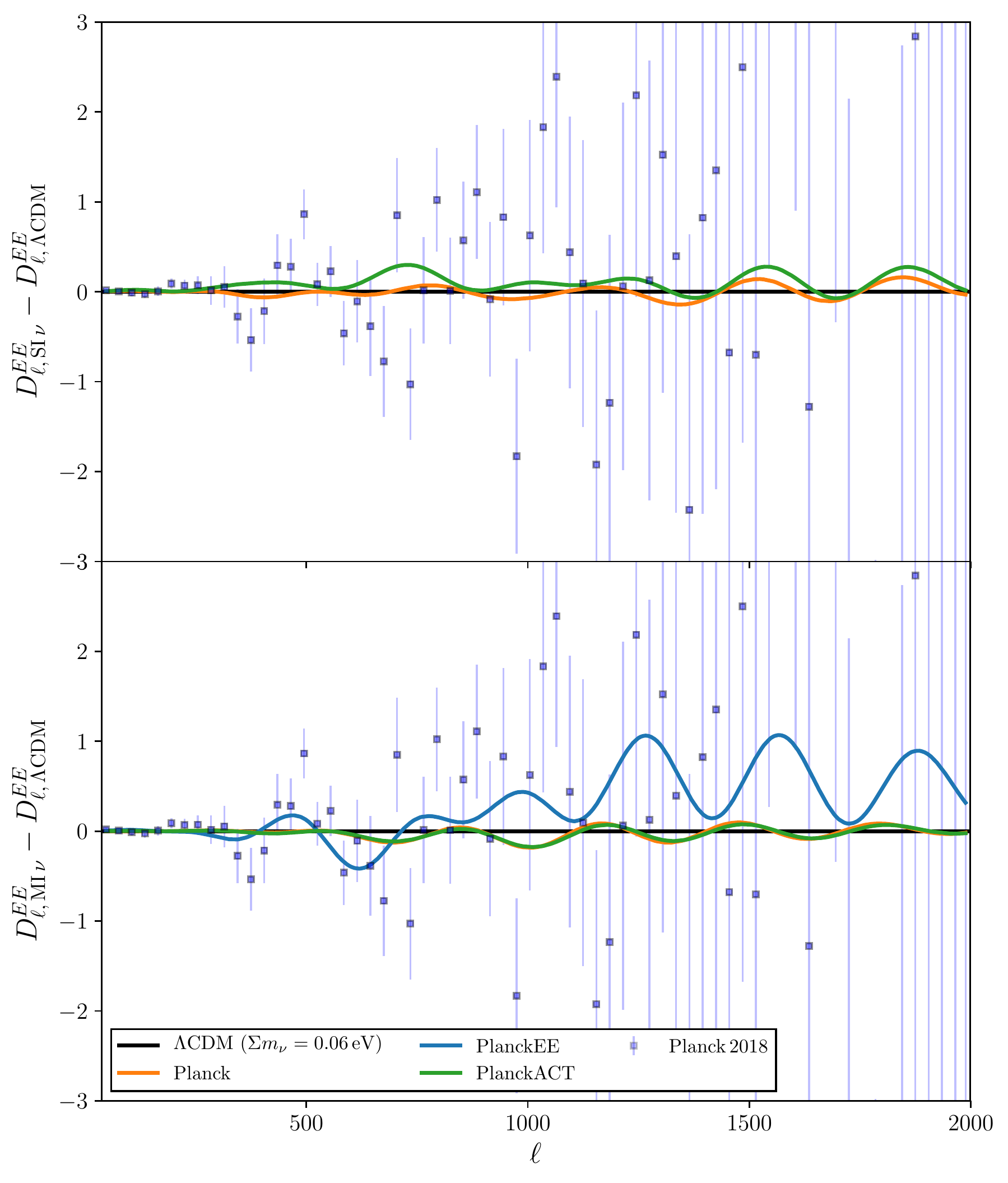}
\caption{Residuals of best-fit EE power spectra relative to $\Lambda \mathrm{CDM}$ (ACT best-fit $\Lambda \mathrm{CDM}$ for the left panel, \textit{Planck} best-fit $\Lambda \mathrm{CDM}$ for the right panel). The upper panels are SI$\nu$ best fits and the lower panels are MI$\nu$ best fits. }
\label{fig:EE}
\end{center}
\end{figure*}


\section{Adding Distance Ladder Measurements}
\label{app:H0me}


For completeness, we now discuss the addition of Cepheid calibrated distance-ladder Hubble constant measurements, $H_0 = 73.24 \pm 1.74$ km/s/Mpc measured by the SH$_0$ES team in 2016~\cite{HST}, to our parameter constraints. Though there are more recent local measurements using Cepheid calibrated SNIa~\cite{Riess:2020fzl, Riess:2021jrx}, TRGB calibrated SNIa~\cite{trgb1,trgb2}, and strong lensing time delay~\cite{tdcosmo}, we utilize the SH$_0$ES team's 2016 measurements to directly parallel the study done in~\cite{kreisch}.
Similar to when adding BAO measurements (see \autoref{sec:BAO_H0}), adding the SH$_0$ES $H_0$ measurements decreases the significance of the MI$\nu$ mode and decreases the width of the SI$\nu$ mode (compare ACT + \textit{WMAP} + BAO + SH$_0$ES to ACT + \textit{WMAP} + BAO in \autoref{fig:posteriors_bao}). Adding in distance ladder $H_0$ measurements from SH$_0$ES nearly eliminates the MI$\nu$ mode as a viable statistical possibility. 

\clearpage

\begin{figure*}[t]
\begin{center}
\includegraphics[width=0.75\textwidth]{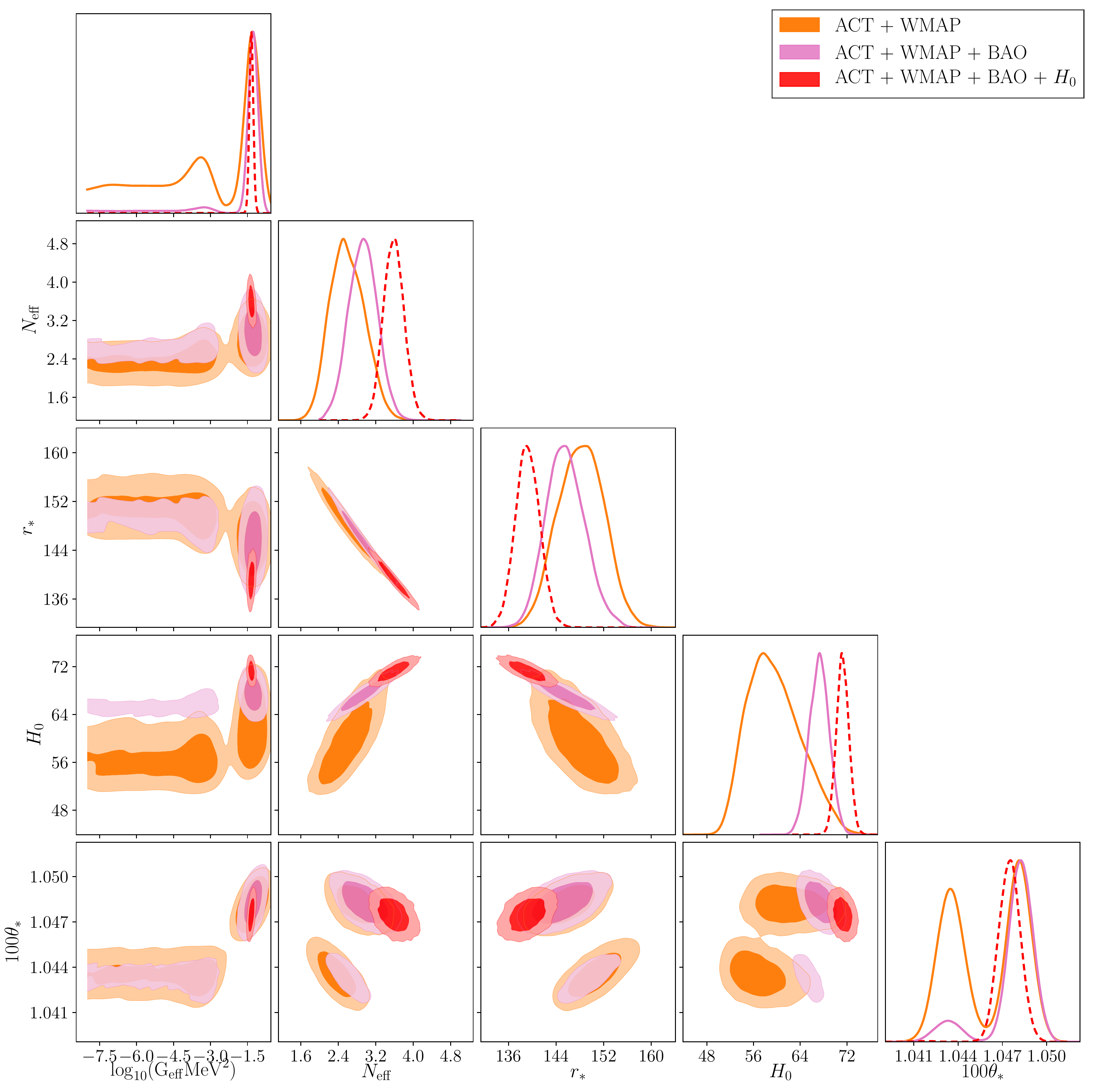}
\caption{Posteriors for \loggeff, $N_{\rm eff}$, $r_*$ (the comoving size of sound horizon at recombination), $H_0$, and $100\theta_*$ (the angular size of sound horizon at last scattering) with ACT+\textit{WMAP} (orange), ACT+\textit{WMAP}+BAO (pink), and ACT+\textit{WMAP}+BAO+SH$_0$ES (red) dataset combinations. The introduction of distance ladder measurements on top of BAO measurements, which already suppresses the MI$\nu$ mode, nearly removes the MI$\nu$ mode.}
\label{fig:posteriors_bao}
\end{center}
\end{figure*}

\section{ACT Wide/ Deep}\label{sec:ACTwidedeep}

\begin{table}[t!]
\begin{minipage}[]{0.48\linewidth}
\includegraphics[width=\textwidth]{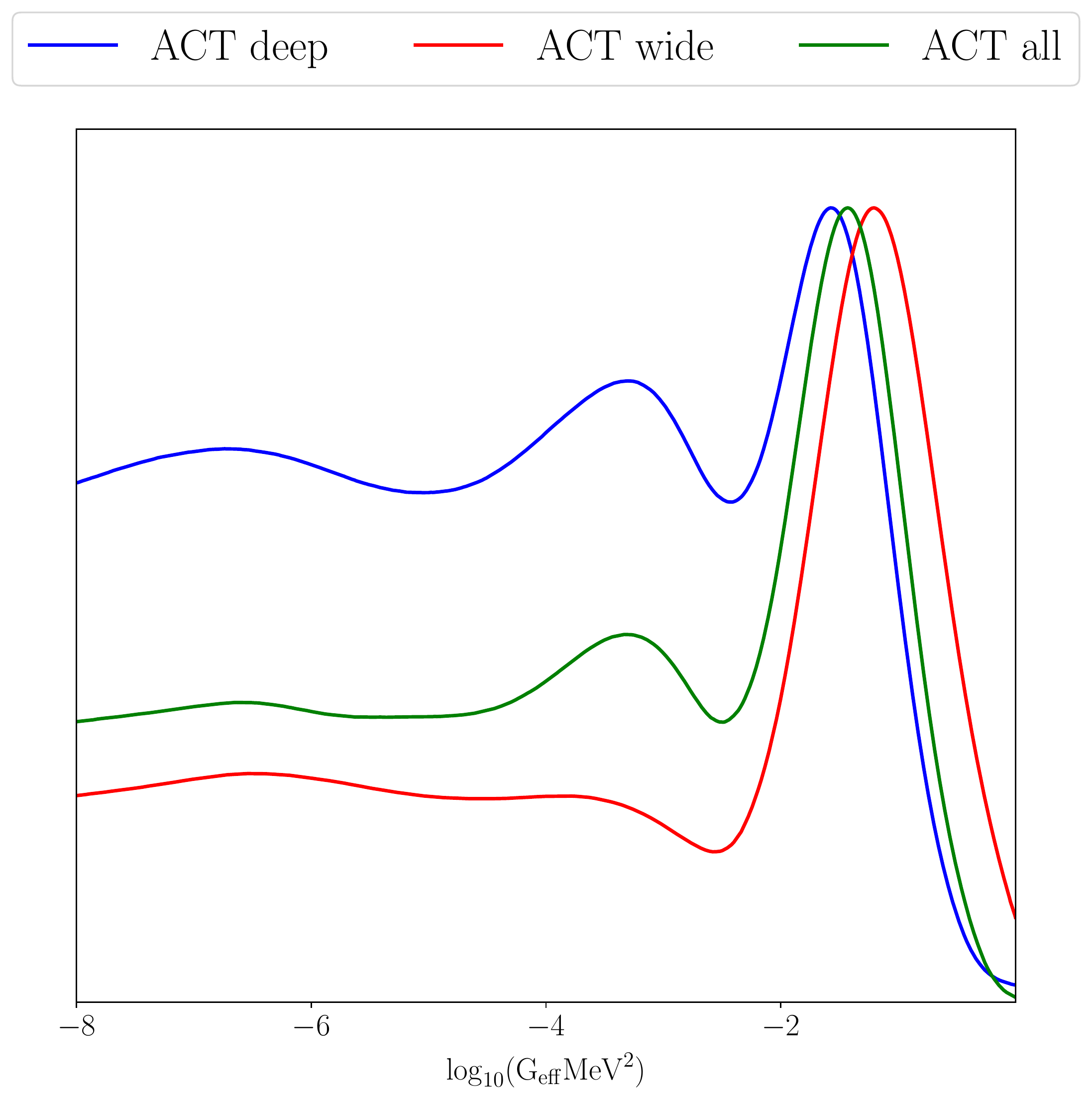}
\captionof{figure}{1D posterior for \geff, obtained by varying the 6 \LCDM\ parameters, \Mnu, \Nnu, and \geff\ with all of ACT (in green), `deep' ACT (in blue), and `wide' ACT (in red). The `wide' only posterior shows the strongest preference for SI$\nu$ but the blue posterior also does slightly prefer SI$\nu$.}
\label{fig:ACTdeepwide}
\end{minipage}
\begin{minipage}[t!]{0.45\linewidth}
\begin{ruledtabular}
\caption{ACT deep and wide comparisons to $\Lambda\mathrm{CDM}$  ($\Delta \chi^2 = \chi^2_{{\rm SI} \nu} - \chi^2_{\rm \Lambda CDM}$). \label{tab:chi2_deepwide}}
\begin{tabular}{ccc} 
Parameter & ACT deep & ACT wide\\ 
\hline \\ [-2ex]
$\Delta \chi^2_{\mathrm{ACT}}$ 	& $ -4.4$	& $-8.0$	 \\
$\Delta \chi^2_{\mathrm{ACT: TT}}$ 	& $-5.0$	& $0.8$ \\
$\Delta \chi^2_{\mathrm{ACT: TE}}$ 	& $0.3$	& $ 1.4$  \\
$\Delta \chi^2_{\mathrm{ACT: EE}}$ 	& $ 0.3$	& 	$ -10.1$\\
\\ [0.5ex] \hline \\[-2ex]
$\Delta \chi^2_{\mathrm{prior}}$ 	& $0.01$	&  $0.01$ \\
$\Delta \chi^2_\mathrm{Total}$ 	& $-4.4$	& $-7.9 $ \\
$\Delta \mathrm{AIC}$ 	& $1.6$	& $-1.9$
\end{tabular}
\end{ruledtabular}
\end{minipage}
\end{table}

\begin{figure*}[t!]
\begin{center}
\includegraphics[width=\textwidth]{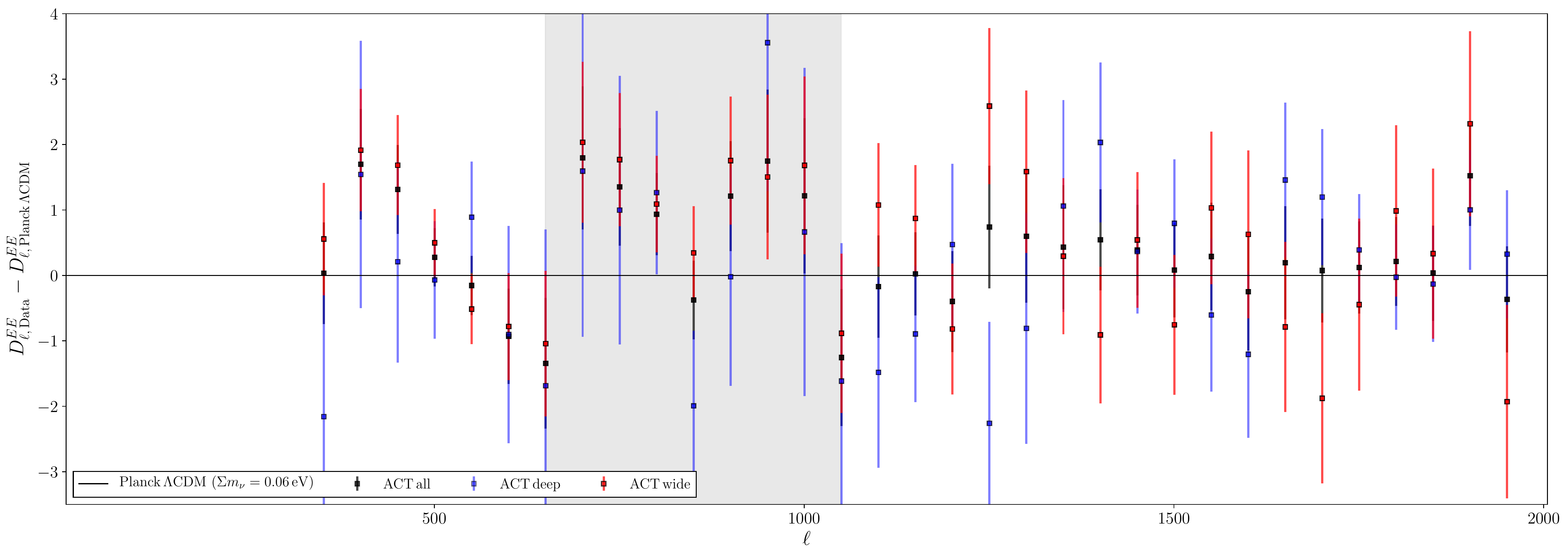}
\caption{Difference for the E-mode polarization data for ACT wide, deep, and total compared to the best-fit $\Lambda$CDM model from \textit{Planck}. In the $700\lesssim \ell \lesssim 1000$ range of interest, the three datasets are not significantly far apart.  }
\label{fig:data_widedeep}
\end{center}
\end{figure*}

\autoref{fig:ACTdeepwide} shows the 1D posterior of \loggeff\ for ACT `deep,' ACT `wide,' and the full ACT dataset. We find that both the posteriors produced with `deep' and `wide' have a preference for \loggeff$\ \sim -1.3$ around which their tallest peaks and best fit models are located. However, the preference for SI$\nu$ over MI$\nu$ is much stronger with `wide' than with `deep,' resulting in the combined ACT posterior which is between the two other posteriors. We show $\Delta\chi^2$ values comparing SI$\nu$ to $\Lambda \mathrm{CDM}$ for the isolated `deep' and `wide' components of ACT in \autoref{tab:chi2_deepwide}. The `wide' E-mode polarization is the only component with a strong enough preference for SI$\nu$ to bring $\Delta {\rm AIC}<0$ compared to \LCDM. 

We show the `wide' and `deep' E-mode polarization residuals in \autoref{fig:data_widedeep}. While the `wide' data show a stronger fluctuation in $700 \lesssim \ell \lesssim 1000$, the large error bars of the `deep' data are still consistent with this fluctuation. Therefore, the discrepancies between the preferences of `wide' and `deep' are within the margins of error of the two components of ACT. 

This is again similar to the findings in Ref.~\cite{colinede} that the feature which the EDE model fits better than \LCDM\ is more pronounced in the `wide' component (specifically `wide' EE) than the `deep,' further suggesting that the two different physical models maybe describing the same feature in ACT data.

\section{Helium Abundance}\label{sec:helium}

\begin{figure*}[b!]
\begin{center}
\includegraphics[width=0.7\textwidth]{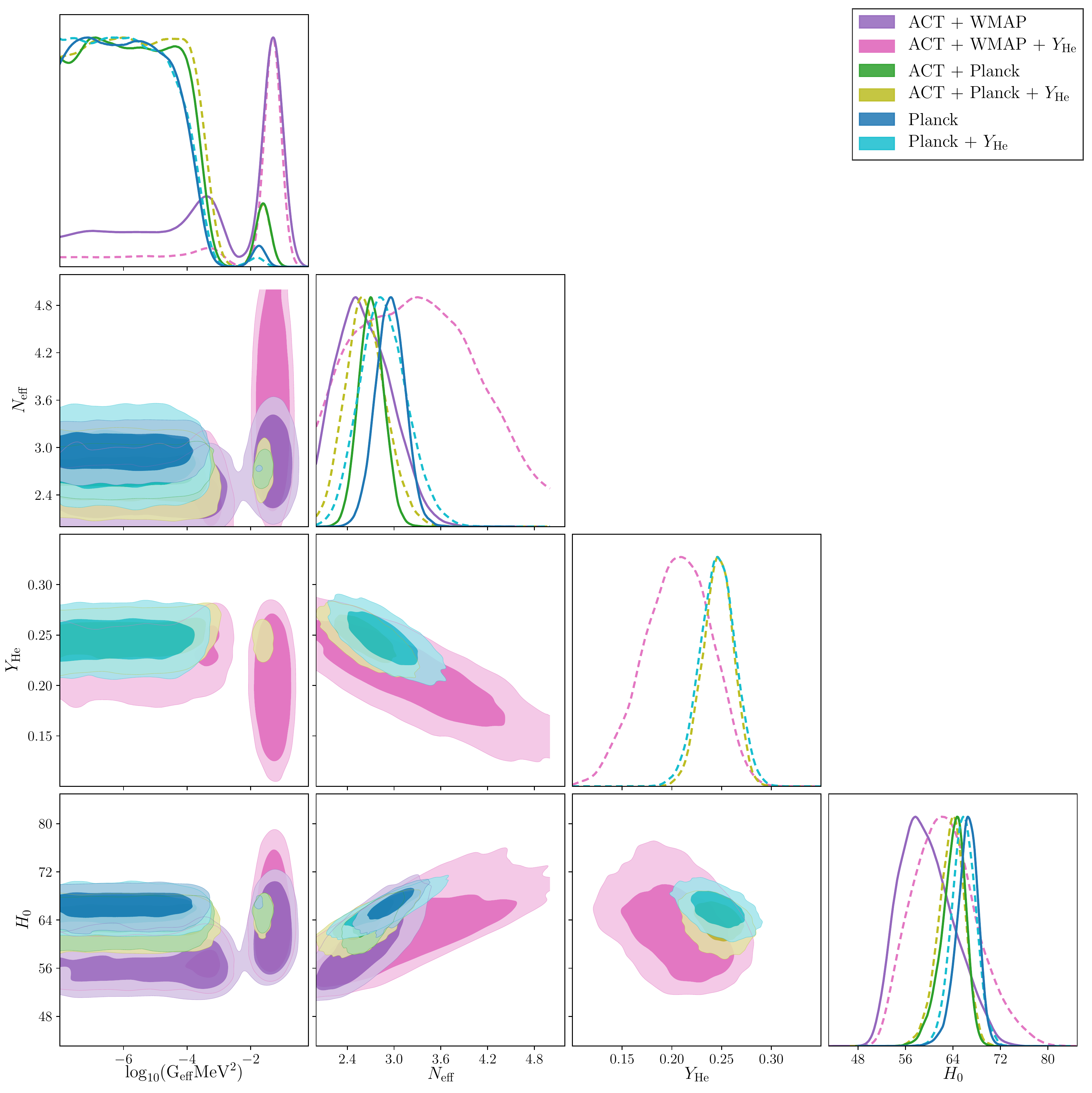}
\caption{Posteriors for \loggeff, $N_{\rm eff}$, $Y_{\rm He}$, and $H_0$ for various CMB dataset combinations. Notice how letting $Y_{\rm He}$ vary decreases the significance of MI$\nu$ with ACT+\textit{WMAP}, but when \textit{Planck} is involved, there is hardly any change as \textit{Planck} is able to constrain $Y_{\rm He}$ without BBN inputs. }
\label{fig:posteriors_yhe}
\end{center}
\end{figure*}

In this section, we consider the impact of freeing the helium abundance $Y_{\rm He}$ away from its Big-Bang Nucleosynthesis (BBN) value. We now thus consider a four-parameter extension of $\Lambda$CDM in which $G_{\rm eff}$, $N_{\rm eff}$, $\sum m_\nu$, and $Y_{\rm He}$ are allowed to vary freely. Our results are summarized in \autoref{fig:posteriors_yhe} for different data set combinations. For ACT + \textit{WMAP}, freeing the helium fraction boosts the SI$\nu$ mode, largely due to the large range of $N_{\rm eff}$ and $Y_{\rm He}$ values that are now accessible within this mode of the posterior. In the Standard Model, BBN predicts a larger helium yield $Y_{\rm He}$ as $N_{\rm eff}$ is increased. However, maintaining the ratio of the photon scattering rate to the Hubble rate constant near recombination, which is necessary to leave the CMB invariant~\cite{Cyr-Racine:2021oal}, requires $Y_{\rm He}$ to \emph{decrease} as  $N_{\rm eff}$ is increased. This can be seen in the strong anticorrelation between  $N_{\rm eff}$ and $Y_{\rm He}$ in \autoref{fig:posteriors_yhe} for the ACT + \textit{WMAP} + $Y_{\rm He}$ case. Since the MI$\nu$ mode is constrained to have low values of $N_{\rm eff}$ as explained above, only the SI$\nu$ mode can exploit this degeneracy once the helium abundance is allowed to vary freely. 

On the other hand, freeing $Y_{\rm He}$ in the presence of \textit{Planck} data has little impact on the posterior, illustrating the ability of \textit{Planck} data to constrain the helium abundance even without BBN inputs. Just as in our baseline model, the \textit{Planck} data strongly disfavor the SI$\nu$ mode compared to the MI$\nu$ mode.

\end{widetext}

\bibliography{nu_ref}  

\clearpage

\end{document}